\documentclass[authoryear,preprint,review,12pt]{elsarticle}

\usepackage{amssymb}
\usepackage{caption}

\usepackage{lipsum}
\usepackage{url}
\usepackage{float}
\usepackage{microtype,siunitx,booktabs}
\usepackage{listings}

\usepackage{longtable} 

\usepackage{glossaries}
\usepackage{hyperref}

\def\aj{AJ}%
\def\apj{ApJ}%
\def\apjs{ApJS}%
\def\aap{A\&A}%
\def\mnras{MNRAS}%
\def\pasp{PASP}%
\journal{Astronomy $\&$ Computing}

\begin{document}

\begin{frontmatter}



\title{Rotation and Flipping Invariant Self-Organizing Maps with Astronomical Images: A Cookbook and Application to the VLA Sky Survey QuickLook Images}


 \author[1]{A.~N. Vantyghem}
\affiliation[1]{organization={University of Manitoba, Department of Physics and Astronomy}, city={Winnipeg}, postcode={R3T 2N2}, state={MB}, country={Canada}}

\author[2,3,4]{T.~J. Galvin}
\affiliation[2]{organization={International Centre for Radio Astronomy Research, Curtin University}, city={Bentley}, postcode={6102}, state={WA}, country={Australia}}
\affiliation[3]{organization={CSIRO Space \& Astronomy}, addressline={PO Box 1130}, 
city={Bentley}, postcode={6102}, state={WA}, country={Australia}}
\affiliation[4]{organization={Western Sydney University}, addressline={Locked Bag 1797}, city={Penrith}, postcode={2751}, state={NSW}, country={Australia}}

\author[1]{B. Sebastian}

\author[1]{C.~P. O'Dea\corref{cor1}}
\ead[C.~P. O'Dea]{odeac@umanitoba.ca}

\author[5,1]{Y.~A. Gordon}
\affiliation[5]{organization={University of Wisconsin-Madison, Department of Physics},addressline={1150 University Ave}, city={Madison},postcode={53706}, state={WI}, country= {USA}}

\author[1]{M. Boyce}

\author[6]{L. Rudnick}
\affiliation[6]{organization={University of Minnesota, Minnesota Institute for Astrophysics, School of Physics and Astronomy}, addressline={116 Church Street SE}, city={Minneapolis}, postcode={55455}, state={MN}, country={USA}}



\author[10]{K. Polsterer}
\affiliation[10]{Astroinformatics, HITS gGmbH, Schloss-Wolfsbrunnenweg 35, 69118 Heidelberg, Germany}

\author[7,11]{Heinz Andernach}
\affiliation[7]{organization={Thueringer Landessternwarte}, addressline={Sternwarte 5},  city={Tautenburg}, postcode={D-07778}, country={Germany}}
\affiliation[11]{organization={Permanent Address:Univ. de Guanajuato, Depto. de Astronomia}, addressline={Callejon de Jalisco s/n}, city={Guanajuato}, postcode={C.P. 36023, GTO}, country={Mexico}}


\author[8]{M. Dionyssiou}
\affiliation[8]{organization={University of Toronto, Dunlap Institute for Astronomy and Astrophysics}, addressline={50 St George Street}, city={Toronto}, postcode={M5S 3H4}, state={ON}, country={Canada}}

\author[8]{P. Venkataraman}

\author[3,4]{R. Norris}

\author[1]{S.~A. Baum}

\author[9]{X. R. Wang}
\affiliation[9]{CSIRO Data61, Australia, PO Box 76, Epping, NSW 1710, Australia}

\author[3]{M. Huynh}

\begin{abstract}
Modern wide field radio surveys typically detect millions of objects. Manual determination of the morphologies is impractical for such a large number of radio sources. Techniques based on machine learning are proving to be useful for classifying large numbers of objects. 
The self-organizing map (SOM) is an unsupervised machine learning algorithm that projects a many-dimensional dataset onto a two- or three-dimensional lattice of neurons. This dimensionality reduction allows the user to visualize common features of the data better and develop algorithms for classifying objects that are not otherwise possible with large datasets. To this aim, we use the PINK 
implementation of a SOM. PINK incorporates rotation and flipping invariance so that the SOM algorithm may be applied to astronomical images. In this cookbook we provide instructions for working with PINK, including preprocessing the input images, training the model, and offering lessons learned through experimentation. 
The problem of imbalanced classes can be improved by careful selection of the training sample and increasing the number of neurons in the SOM (chosen by the user). Because PINK is not scale-invariant, structure can be smeared in the neurons. This can also be improved by increasing the number of neurons in the SOM.

We also introduce \textsc{pyink}, a Python package used to read and write PINK binary files, assist in common preprocessing operations, perform standard analyses, visualize the SOM and preprocessed images, and create image-based annotations using a graphical interface. A tutorial is also provided to guide the user through the entire process. We present an application of PINK to VLA Sky Survey (VLASS) images. We demonstrate that the PINK is generally able to group VLASS sources with similar morphology together. We use the results of PINK to estimate the probability that a given source in the VLASS QuickLook Catalogue is actually due to sidelobe contamination.
\end{abstract}



\begin{keyword}
astronomy software \sep software documentation \sep astronomy image analysis
\end{keyword}

\end{frontmatter}




\section{Introduction}
\label{introduction}
\label{sec:intro}
Stimulated by new large radio surveys  that generate catalogs of millions of sources,  a major challenge is to develop machine-learning algorithms for the classification of radio sources. These machine learning algorithms come in two varieties: supervised and unsupervised. Supervised algorithms need to be trained on a sample of sources that have already been classified by eye, typically based on the results of a citizen science project such as Radio Galaxy Zoo
 \citep{Banfield2015}. The most successful supervised algorithms for this purpose are convolutional neural nets \citep[e.g.][]{Scaife2021, Wu2019, Alger2018, Aniyan2017}, but they all suffer from the challenge of needing large manually-generated training sets. Unsupervised algorithms  
\citep[e.g.][]{Ghahramani2004} instead identify patterns within the data without the need for user-supplied training sets, resulting in a small number of classes that can then be manually labelled.

\subsection{Self-Organizing Maps}



The self-organizing map \citep[SOM;][]{Kohonen1988, Kohonen2001} is an unsupervised neural network that projects high-dimensional data onto a low-dimensional ``map'' (or ``lattice''). The map is generally two or three-dimensional, consisting of a rectangular or hexagonal grid of ``neurons'' (also known as ``nodes'' or ``prototypes''). The neurons have fixed positions on the lattice, but are associated with a vector of weights in the same shape (i.e., dimensionality) as the data (i.e., a 1-dimensional vector or 2-dimensional image) that are updated during training to reflect dominant features in the input data.

An important feature of the SOM is that the neurons are arranged coherently. Neurons that are in close proximity on the map have more similar morphologies than neurons that are far apart from each other. This similarity is measured through a distance metric, where a high similarity corresponds to a small distance. The Euclidean distance ($\Delta$) is a common choice. It is defined as
\begin{equation}
    \Delta(A, B) = \sqrt{\Sigma_i (A_i - B_i)^2},
\end{equation}
where $A$ and $B$ correspond to a particular neuron and image and the summation is over all pixels $i$ in those images.

In many astronomical studies, the orientation of an object on the sky is irrelevant. In these cases, the SOM must be modified to incorporate invariance to both rotation and a  flip. 
The package used throughout this cookbook is PINK\footnotemark
(Parallelised rotation and flipping INvariant Kohonen map, see  \citet{PINK})
 which accomplishes this by brute force, generating a series of rotations and flips to each input image. The image transformation that is best matched to one of the neurons is used for that object.
\footnotetext{Available: \href{https://github.com/HITS-AIN/PINK}{https://github.com/HITS-AIN/PINK}}

\begin{figure}
    \centering
    \includegraphics[width=0.99\textwidth]{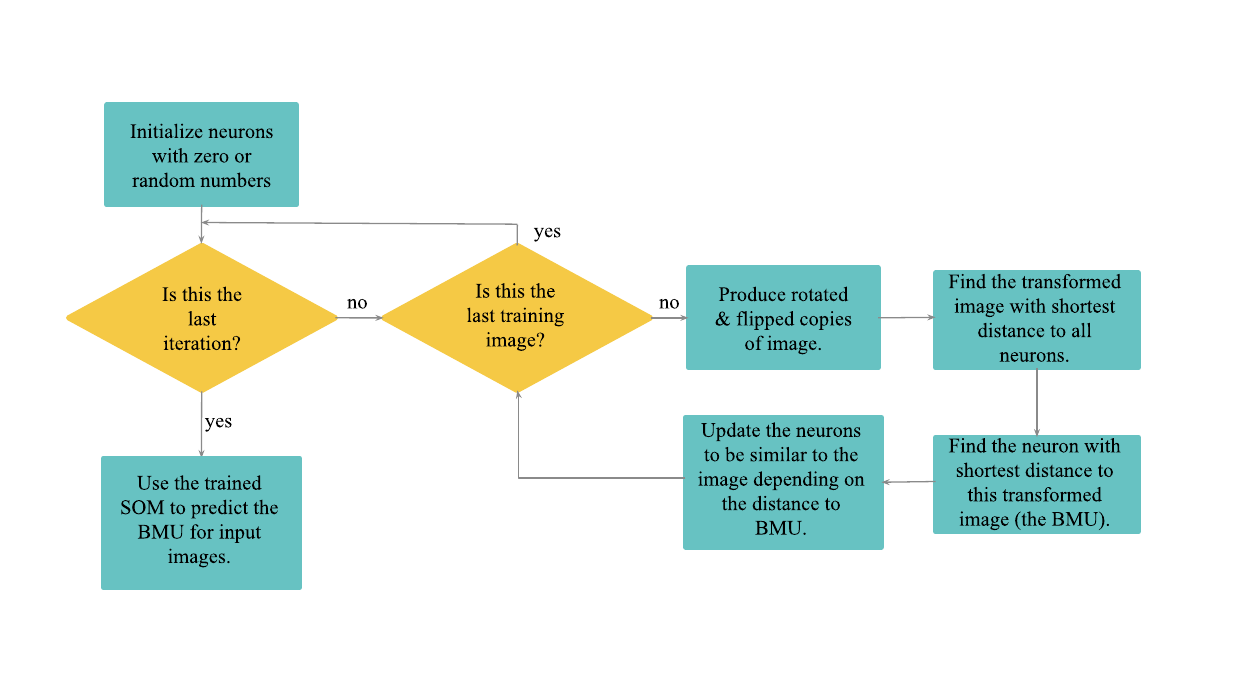}
    \caption{Flowchart illustrating the algorithm of Self-Organizing map training}
    \label{fig:flowchart}
\end{figure}

The algorithm for training a rotationally invariant SOM is outlined in the flowchart in Figure \ref{fig:flowchart} and is described as follows.
\begin{enumerate}
    \item 
    Initialize the weights for each neuron. For a 1-dimensional vector these correspond to feature values, while in a 2-dimensional image these are pixel values. The starting values can be set to zero or filled with randomly generated numbers.
    \item For each entry in the training sample:
    \begin{enumerate}
        \item Create the rotated and flipped copies of the input image.
        \item Calculate the similarity (Euclidean distance) between all transformed input images and all neurons.
        \item For each neuron, find the transformed image that results in the lowest Euclidean distance. If multiple image transformations result in the same distance, one is chosen at random.
        \item Identify the neuron with the lowest Euclidean distance from amongst the transformed images selected in the previous step. This neuron is known as the best-matching unit (BMU), though we refer to it as the best-matching neuron.
        \item Update the pixel values of each neuron so that they are more similar to the input image. Neurons in close proximity to the best-matching neuron are updated more significantly than distant ones. This is controlled through the neighbourhood function (also known as a distribution function) $r(n_1, n_2)$, where $n_1$ and $n_2$ are the lattice coordinates of two neurons in the SOM. A Gaussian is commonly used as a neighbourhood function,
        \begin{equation}
            r(n_1, n_2) = \frac{1}{\sqrt{2\pi \sigma^2}}\exp{\left[ \frac{-s(n_1, n_2)}{2\sigma^2} \right]}.
        \end{equation}
        Here $s$ is the separation between two neurons on the lattice, which can account for cyclical continuity at the SOM edges. $\sigma$ controls the size of the neighbourhood and can be updated between steps.
        
        The neuron weights are then updated according to
        \begin{equation}
            w_i' = w_i + \delta (D-w_i) \times r(n_i, BMU),
        \end{equation}
        where $w_i$ and $w_i'$ are the initial and final weights of the $i$'th neuron, $D$ is the transformed image. The learning rate, $\delta$, is a scalar that controls the magnitude of the adjustment.
    \end{enumerate}
    \item Repeat for a fixed number of iterations.
    \item Using the trained SOM, measure the distance between each input image and every neuron in order to determine the best-matching neuron for each image.
\end{enumerate}


\subsection{SOM uses in astronomy}

As the SOM is a powerful tool for clustering and dimensionality reduction, it is well-posed to handle the large data volumes in astronomy. 
It has been applied successfully to the calibration of photometric redshifts \citep{Carrasco2014, Masters2015, Wright2020}, the classification of variable stars \citep{Brett2004, Armstrong2016}, gamma-ray bursts \citep{Rajaniemi2002}, and galaxy morphology \citep{Naim1997}, and has been used to identify outlying objects from quasar samples \citep{Meusinger2012}. These studies all train the SOM on a list of 1-dimensional features, such as photometric colours, morphological heuristics, or a spectrum.

The image-based rotationally invariant SOM has been adopted by several studies that aim to classify radio galaxy morphologies. 
\citet{Galvin2019} attempted to reproduce citizen science morphological classifications of FIRST sources from Radio Galaxy Zoo \citep{Banfield2015}. They trained a SOM containing channels for both FIRST and WISE images in order to identify radio galaxy morphologies that include the connection with infrared (IR) wavelengths. A random forest classifier was then used to match the Radio Galaxy Zoo labels, which describe a radio source in terms of its number of distinct islands and total number of local maxima. The Euclidean distances between an image and every neuron on the SOM were used in this analysis.

\citet{Galvin2020} employed a SOM as part of an algorithm to combine radio components into complex sources without the need for training labels. This was accomplished by training a SOM on FIRST and WISE images to produce a set of object morphologies. As this provides a manageable number of neurons (1600), masks were manually created for each one in order to encapsulate the radio components belonging to the same source as the central component and identify the likely position of the IR host. When an image is matched to a neuron, the corresponding masks (or ``filters'', see Section \ref{sec:image-antns}) are transformed to the image frame and used to identify the radio components that are part of the same source. \citet{Galvin2020} used this technique to identify both resolved radio components with a single IR host as well as a set of 17 giant radio galaxies between 700 and 1100 kpc in size.

\citet{Mostert2021} used a SOM to help identify radio morphologies in the Low Frequency Array (LOFAR, \citep{Shimwell2017}) Two-metre Sky Survey (LoTSS). The $10\times 10$ SOM with cyclic boundary conditions was then combined with the accompanying LoTSS morphological labels in order to associate each neuron with its most common morphology. Most neurons consist of mainly one or two of the six morphological classes, while two of the 100 neurons have a mix of morphologies. They further identified the 100 sources with the rarest morphologies based on their minimum Euclidean distance over all neurons.






\subsection{A practical overview for the SOM}

While the applications for the rotation and flipping invariant SOM are numerous, several key steps are common to the process.
First, a training sample must be defined and the corresponding image cutouts must be obtained and preprocessed.
The preprocessed images are then used to train the SOM. Any sample -- the training sample or a new one preprocessed in the same way as the training data -- can then be mapped onto the SOM to measure the similarity between each image and each neuron as well as the transform parameters required to align the two.

With this information in hand, a multitude of analyses can be conducted. 
The first step is often to annotate the neurons. These can include, but are not limited to, a simple text label (e.g. a morphological classification) or a labelled mask identifying various features within the neuron (see Section \ref{sec:image-antns}). Each catalog entry inherits the annotations from its best-matching neuron.

The simplest experiment is to train a SOM on the desired dataset, label each neuron with a single text label, map all images onto the SOM and identify their best-matching neuron, and finally apply the label associated with the best-matching neuron to the associated images.

\subsection{Overview}

This cookbook is structured as follows. In Section \ref{sec:software} we introduce the required software. Section \ref{sec:prepro} provides guidelines for preprocessing images and outlines utility functions for this purpose. Instructions and recommendations for training a SOM is provided in Section \ref{sec:training}. Section \ref{sec:mapping} demonstrates how to work with the binary files produced through the training process. In Section \ref{sec:annotations} we discuss various methods for annotating each neuron in the SOM, both with text labels and image-based masks. The hardware requirements for training a SOM are presented in Section \ref{sec:hardware}. 
In Section \ref{sec:sidelobes}, we discuss the use of the SOM to find sidelobes in images of the Very Large Array Sky Survey \citep[VLASS,][]{Lacy2020}.
Finally, in Section \ref{sec:lessons} we discuss several lessons that were learned through experimentation with the SOM.

Throughout this cookbook we make reference to an example experiment which had the goal of using a SOM to group radio components into sources. 
The starting point for this work is the  catalog of radio components from VLASS QuickLook (QL, minimally processed first look) epoch 1 (QL1)
\citep{Gordon2021}, which each represent individual blobs of emission that are sometimes a part of larger, more complex structures. The SOM is trained on preprocessed images with two channels -- one for the VLASS cutout centered on the corresponding catalog entry, and an IR cutout of the same field of view from the 
unblurred images from the Wide field Infrared Survey Explorer telescope \citep[unWISE,][]{Wright2010,  Lang2014}. 
We then used image-based annotations to dictate which radio emission in each neuron should be grouped together as well as the likely location of the IR host by utilising the unWISE source catalog \citep{Schlafly2019}.

Before continuing we make a brief comment on terminology. Each row in an input catalog that is to be processed through the SOM is referred to as a catalog entry. An image {cutout} is an image from a small region centered on a catalog entry. A preprocessed {image} consists of one or more image cutouts for a single entry that have been scaled for use in the SOM. Following this convention, an image is one entry from a PINK image binary (see Sections \ref{sec:binaries} and \ref{sec:prepro}).

\section{Software}
\label{sec:software}

Two pieces of software are discussed throughout this work. PINK is used to train the SOM and map all preprocessed images onto it. The Python code \textsc{pyink}\footnote{\url{https://github.com/tjgalvin/pyink}} provides a convenient wrapper for the PINK binary files, includes utilities for using and analyzing the SOM, and provides an interface for image-based annotations (see Section \ref{sec:image-antns}).

\subsection{The PINK Binary Files}
\label{sec:binaries}

PINK uses a series of binary files to encode the data it uses\footnote{\href{https://github.com/HITS-AIN/PINK/blob/master/FILE_FORMATS.md}{PINK binary formats}}. These come in four types: the image binary, SOM, mapping, and transform.
The image binary is the data file used to either train a SOM or be mapped onto one. The SOM binary records the shape of the SOM and the data contained in each of its neurons. The mapping binary is the result of mapping an image binary onto a SOM. It contains a similarity measure relating each entry in the image binary to every neuron in the SOM. Finally, the transform binary records the rotation angle and whether a flip is required to best match an image to the corresponding neuron. 





Each binary begins with an integer specifying the file format version. This refers to the version of PINK, which is currently 2. The file formatting changed between versions 1 and 2. All information presented here is for version 2.

The data type is an integer corresponding to the typing used for the data. The options are listed in Table \ref{tab:data_type}. Currently, however, only 32-bit floating point numbers are supported.

\begin{table}[]
    \centering
    \begin{tabular}{c c}
    PINK Integer & Data Type \\
    \hline
        0 & float 32 \\
        1 & float 64 \\
        2 & integer 8 \\
        3 & integer 16 \\
        4 & integer 32 \\
        5 & integer 64 \\
        6 & unsigned integer 8 \\
        7 & unsigned integer 16 \\
        8 & unsigned integer 32 \\
        9 & unsigned integer 64 \\
    \end{tabular}
    \caption{The data types that can be provided to PINK binary files.}
    \label{tab:data_type}
\end{table}

The data layout is a sequence of numbers specifying the shape of the data. The first digit is either a 0 or 1, indicating whether the data is arranged in a Cartesian (0) or hexagonal (1) layout. The following digits indicate the number of dimensions and then the size of each axis. Cartesian layouts support either 2 or 3 dimensions, but hexagonal layouts are always 2-dimensional. For example, the data layout for 3-dimensional Cartesian data with 2 stacked images each 128$\times$128 pixels in size is ``0 3 2 128 128''.

The SOM and neuron layouts are analogous to the data layout -- the SOM layout is either Cartesian (0) or hexagonal (1), and the neuron layout is the number of dimensions of each neuron followed by the size of each dimension.

\section{Preprocessing}
\label{sec:prepro}

\subsection{Preprocessing Guidelines}
\begin{figure}
    \centering
    \includegraphics[width=0.99\textwidth]{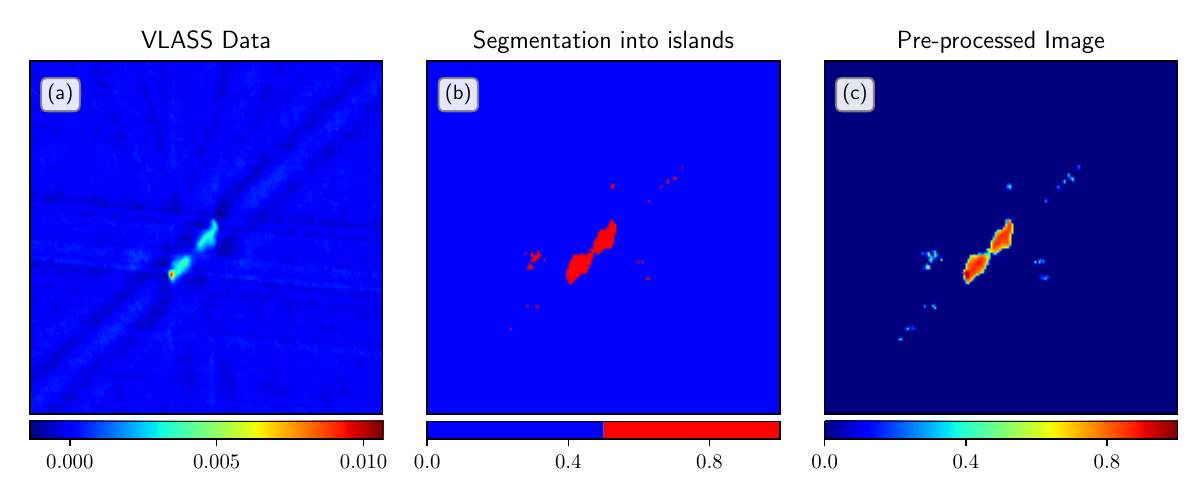}
    \caption{Example of preprocessing on a VLASS image cutout. (a) raw image of a VLASS double source (b) segmented islands identified in the cutout region (c) final preprocessed image normalized on a scale of 0 to 1. }
    \label{fig:preprocess}
\end{figure}
In the preprocessing stage a \textsc{PINK} image binary is created from scaled image cutouts for each entry in the catalog of interest. Multiple cutouts (channels) can be included for each catalog entry. This can represent, for example, the same field of view at different wavelengths. Every cutout must have the same number of pixels.

Steps in the  preprocessing of a VLASS image are shown in Figure \ref{fig:preprocess}.
The basic workflow for preprocessing is as follows. 
The exact steps depend on the science case.
\begin{enumerate}
    \item Mask unrelated structures in the field of view.
    \item Apply a flux threshold to mask the noise.
    \item Segment the image cutout into islands (optional).
    \item Apply a transfer function to the image cutout (or islands), normalizing on a 0 to 1 scale.
    \item Apply a weighting to each channel based on the sum of all pixel brightnesses.
\end{enumerate}

The similarity measure used by the SOM necessitates that all images be normalized {  in brightness. } The simplest is to use a 0 to 1 scaling with any transfer function deemed appropriate. It can be effective to first segment each image into isolated islands and normalize each independently, as this allows the SOM to prioritize the positioning of the islands as opposed to their relative brightness. Thus, 
island segmentation returns a collection of binary masks. These are applied separately to the data, and the minmax function is then applied. The result is a collection of islands/clumps that are minmax normalized.

Each channel must also be supplied with an appropriate weighting. This will effectively dictate the priority that the SOM assigns to patterns in each channel; a large relative weighting in one channel will prioritize the spatial patterns in that channel over the others. Along these lines, it is critical to consider the sum of all pixel brightnesses in each channel in order to set the appropriate weighting.  For example, if an image consists of infrared and radio channels, and the collective brightness of all pixels in the radio channel surpasses that of the IR channel, this would result in an overweighting of the features in the radio channel.

The SOM is highly effective at identifying patterns in the underlying noise, such as the sidelobe patterns of radio interferometers. If this information is not useful, it is best to mask it out.
Similarly, if objects unrelated to the source of interest are present in the field of view, the SOM may identify patterns based on their positioning. This can be addressed by adjusting the channel weight or masking out the objects unrelated to the main source.

\subsection{Creating an Image Binary (ImageWriter)}

The PINK image binary is a series of numbers that encode header-style information followed by the data. The formatting is as follows.
{\footnotesize
\begin{verbatim}
  <file format version> 0 <data-type> <number of 
    entries> <data layout> <data>
\end{verbatim}}

A data type of 0 is used to specify float32. The number of entries is the number of catalog rows that will be included in the binary. 
The data layout begins with a 0 or 1 to indicate whether the data is Cartesian (0) or hexagonal (1). The following digits indicate the number of dimensions and the sizes of each. For example, a data layout of "0 2 150 150" refers to two-dimensional Cartesian data, each containing $150\times 150$ pixels.

The creation of the image binary and the details of the header digits are handled automatically by the ImageWriter class.\footnote{ Note that, at this time, hexagonal layouts are supported by PINK, but not by the Python wrapper code.}
An image binary can be initialized as follows.

{\footnotesize
\begin{verbatim}
ImageWriter(  # Create an Image Binary file.
    binary_path,   # Output binary file name
    data_layout,   # Cartesian (0) or hexagonal (1) layout
    data_header,   # Image dimensions, e.g. (2, 100, 100)
                   # for two channels with 100x100 images
    comment=None,  # A comment to place in the file header
    clobber=False, # Overwrite an existing file
)
\end{verbatim}}

Once the ImageWriter instance has been created, images can be added to it sequentially.

{\footnotesize
\begin{verbatim}
ImageWriter.add(
    img, # Image array of shape (z, y, x). z corresponds
         # to the number of channels. If z = 1 it may be
         # ignored.
    nonfinite_check = True, # Ensure that only finite 
                            # values are in images
    attributes = None,      # Information to store in the
                            # `records` attribute
)
\end{verbatim}}

For example, given a list of $100\times 100$ pixel images that have been preprocessed (\verb|preprocessed_images|), the image binary can be populated as follows. Here we also include the entry index using the attributes keyword, which will create a records file that is automatically read in with the image binary.
%
{\footnotesize
\begin{verbatim}
import pyink as pu
imbin = pu.ImageWriter(``example_binary.bin'',0,(100,100))
# Cartesian layout. Single channel containing 100x100
# images.

for i, img in enumerate(preprocessed_images):
    imbin.add(img, attributes=i)
\end{verbatim}}

\subsection{Using an Image Binary (ImageReader)}

To load an existing image binary, use the ImageReader class.

{\footnotesize
\begin{verbatim}
ImageReader(
    path,            # Path to the PINK image binary
    record_path=None # Path to a serialised list to
                     # accompany the image binary
                     # Default: path_name.records.pkl
)
\end{verbatim}}

\verb|ImageReader.data| can be used to access the preprocesed images.
This is an array of shape ($N$, $N_C$, $p_y$, $p_x$), where $N$ is the number of images, $N_C$ is the number of channels (e.g. the number of wavelengths used in the image stack), and $p_i$ refers to the number of pixels along axis $i$. The number of image dimensions and shape of an individual image can also be accessed via the \verb|ImageReader.img_rank| and \verb|ImageReader.img_shape| properties.

\subsubsection{Transform Images to Neuron Frame}

Mapping the images onto the SOM involves applying a transformation that includes a rotation and, if necessary, a horizontal flip. The specific axis used for flipping an image is not critical since a flip about one axis can be reproduced by a combination of flips and rotations around any other axis.
As the image binary contains the original images, the transformation must be applied in order to compare them to their best-matching neuron. This can be accomplished using either the \verb|ImageReader.transform| and \verb|ImageReader.transform_images| functions. The former operates on a single image, while the latter can transform multiple images. The transform information can be obtained from the \verb|Transform| binary file (see Section \ref{sec:binaries}).

{\footnotesize
\begin{verbatim}
ImageReader.transform(
    idx,      # The index in the image binary to be 
              # transformed
    transform # A list containing the rotation angle and 
              # a number (0 or 1) indicating whether the
              # image should be horizontally flipped
)
Returns:
    np.ndarray -- The transformed image
\end{verbatim}}

{\footnotesize
\begin{verbatim}
ImageReader.transform_images(
    idx,        # The list of indices in the image binary
                # to be transformed
    transforms  # A list of the transforms to be applied
                # to each index
)
Returns:
    np.ndarray -- Set of transformed images
\end{verbatim}}

\subsubsection{Modifying the Channel Weights}

In an image binary that contains multiple channels, each channel is given a relative weight to influence the patterns learned by the SOM. Should these weights need to be changed, a new image binary can be created from the initial one using the \verb|ImageReader.reweight| function.

{\footnotesize
\begin{verbatim}
ImageReader.reweight(
    binary_path,    # Output path for the new image binary
    old_weights,    # A list containing the weights to all
                    # channels in the original image 
                    # binary
    new_weights,    # A list containing the new weights to
                    # apply to each channel
    verbose=False,  # Print a status update every 1000 
                    # iterations
)
Returns:
    pyink.ImageReader -- New ImageReader instance with 
                         the updated weights.
\end{verbatim}}

\subsection{Preprocessing Tools}

Here we list the tools provided to assist in the preprocessing stage.

\subsubsection{Background Estimation}

The \verb|rms_estimate| function provides multiple methods for estimating the noise level in an image. It uses either the standard deviation or median absolute deviation to flag pixels beyond a specified threshold. The remaining values are then binned and fitted by a second-order polynomial that approximates a Gaussian with significantly lower computational complexity. The fitted parameters are then converted to a standard deviation for the measure of background noise. 

{\footnotesize
\begin{verbatim}
rms_estimate(
    data,              # The data from which the
                       # background is to be estimated.
    mode = `std',      # Clipping mode used to flag 
                       # outlying pixels, either made 
                       # on the median absolute deviation
                       # (`mad') or standard deviation
                       # (`std')
    clip_rounds = 2,   # Number of times to perform the 
                       # clipping of outlying pixels
    bin_perc = 0.25,   # Minimum fraction of bins
                       # required in the fitting 
                       # procedure.
    outlier_thres = 3, # Number of units of the adopted 
                       # outlier statistic required for
                       # an item to be considered an 
                       # outlier.
    nan_check = True   # Remove non-finite values from 
                       # the data
)
Returns:
    float -- Estimated RMS of the supplied image
\end{verbatim}}

\subsubsection{minmax}

The \verb|minmax| function normalizes the data on a 0 to 1 scale. A transfer function should be applied in a separate, earlier step. A mask can be provided to \verb|minmax| to isolate the data to be included in the normalization. This can be used to, for example, normalize each flux island within the image on its own 0 to 1 scale.

{\footnotesize
\begin{verbatim}
minmax(
    data,         # The data to be normalized
    mask = None,  # A boolean mask of the same shape as
                  # the data. Specifies the data to be 
                  # included in the calculation.
)
Returns:
    np.ndarray -- Scaled data
\end{verbatim}}

\subsubsection{Mask the Inner Regions of the Image}

The \verb|square_mask| and \verb|circular_mask| functions return a Boolean array that demarcate either a square or circular region of a specified size centered on the image center.

{\footnotesize
\begin{verbatim}
square_mask(     # Return a boolean array with the inner 
                 # region marked as valid
    data,
    size=None,   # Size of the inner valid region
    scale=None,  # Compute the size of the region in 
                 # proportion to the data shape
)
Returns: 
    np.ndarray -- Square boolean array mask
\end{verbatim}}

{\footnotesize
\begin{verbatim}
circular_mask(   # Return a circular boolean array with 
                 # inner region marked as valid.
    data,
    radius=None, # Radius of the inner valid region
    scale=None,  # Compute the size of the region in 
                 # proportion to the data shape
)
Returns:
    np.ndarray -- Boolean array with a circular valid 
                  region at the center
\end{verbatim}}

\subsubsection{Island Segmentation}

The \verb|island_segmentation| function divides the flux in an image into a set of unique islands. This is based on the \textsc{scikit-image} package\footnote{\href{https://scikit-image.org/docs/dev/api/skimage.morphology.html\#skimage.morphology.closing}{skimage.morphology.closing}}. The threshold specifies the minimum flux required to create an island. The minimum island size can also be specified, removing islands that only occupy a few pixels.

{\footnotesize
\begin{verbatim}
island_segmentation(  # Yield a set of masks that denote 
                      # unique islands in avimage after 
                      # a threshold operation has been 
                      # applied.
    data,
    threshold,  # Threshold level to create the set of
                # islands.
    return_background=False, # Return the zero-index 
                             # background region 
                             # determined by scikit-img.
    minimum_island_size=5,   # Minimum number of pixels 
                             # required for an island 
                             # to be returned.
)
Returns:
    Iterator[np.ndarray] -- Set of island masks
\end{verbatim}}

\subsubsection{Convex Hull}

A convex hull is the smallest convex envelope that encloses a shape. It is analogous to tightening an elastic band around an ensemble of pegs. The \verb|convex_hull| function is based on the \textsc{scikit-image} implementation\footnote{\href{https://scikit-image.org/docs/dev/auto_examples/edges/plot\_convex\_hull.html}{skimage.morphology.convex\_hull\_image}}, but includes a flux threshold to influence the outer extent of the convex hull.

The \verb|convex_hull_smoothed| function determines the convex hull and then smooths it with a Gaussian kernel.

{\footnotesize
\begin{verbatim}
convex_hull(  # Compute the convex hull for an image.
    data,
    threshold=0.01  # Minimum value of the data to 
                    # include in the convex hull.
)
Returns:
    np.ndarray -- 2D Boolean array of the convex hull.
\end{verbatim}}

{\footnotesize
\begin{verbatim}
convex_hull_smoothed(  # Compute and smooth the convex 
                       # hull for an image.
    data,
    sigma,          # Standard deviation of the gaussian 
                    # kernel for smoothing.
    threshold=0.01  # Minimum value of the data to 
                    # include in the convex hull.
)
Returns:
    np.ndarray -- 2D Boolean array for the smoothed
                  convex hull.
\end{verbatim}}

\section{Training}
\label{sec:training}

\begin{figure}
    \centering
    \includegraphics[width=0.99\textwidth]{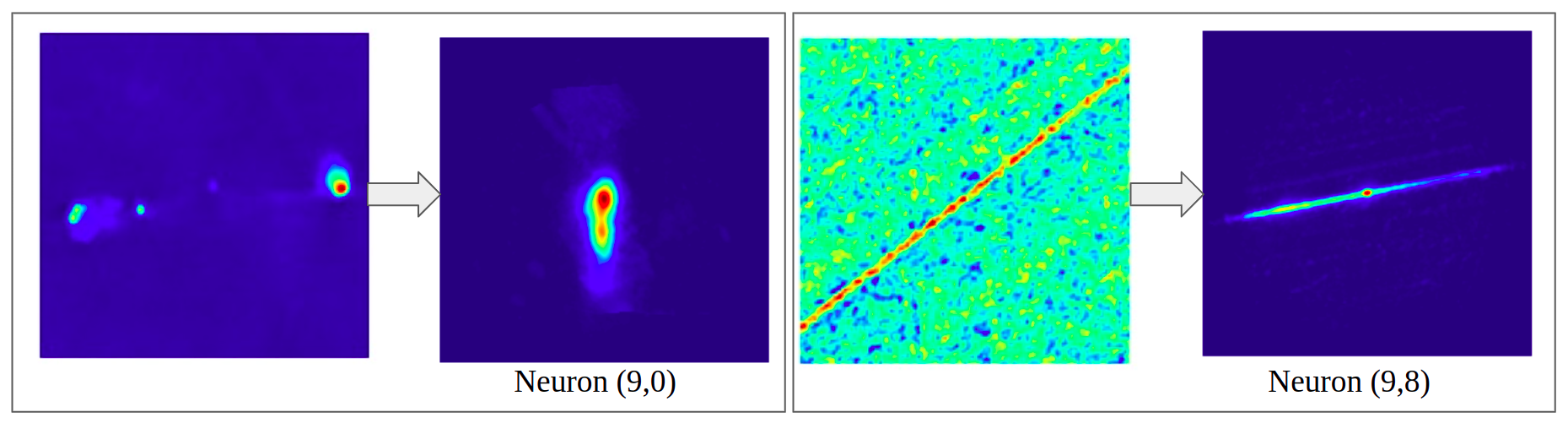}
    \caption{Example of mapping an input VLASS image cutout to a neuron in the SOM. The right image in both the left and right panels shows the neuron, whereas the left image shows the real data. The real images undergo various rotational transformations before being matched with the BMU (see Figure~\ref{fig:transform} for an example). }
    \label{fig:mapping}
\end{figure}

The training of a SOM is performed using \textsc{PINK}. The process requires only an image binary, which can be created using \textsc{pyink} (see Section \ref{sec:prepro}), and the name for the output SOM binary file. In addition, a number of optional command-line arguments can be specified. An exhaustive list is provided in Table \ref{tab:pink}. The syntax is as follows.
{\footnotesize
\begin{verbatim}
    Pink [Options] --train <image-file> <SOM-file>
\end{verbatim}}
Mapping an image binary onto a SOM is done in a similar way.
{\footnotesize
\begin{verbatim}
    Pink [Options] --map <image-file> <map-file> <SOM-file>
\end{verbatim}}
Here ``map file'' is the output of this stage. It is also recommended to use \verb|--store-rot -flip <transform-name>| to obtain the transform binary.

The optional arguments include a number of parameters that customize both the SOM and the training process. Tuning these parameters is necessary to obtain the best possible SOM for the research interest in question. The main parameters that require tuning are the SOM dimensions (\verb|som-width| and \verb|som-height|) and the distribution function (\verb|dist-func|).

Three distribution functions, also known as neighbourhood functions, are currently supported in \textsc{PINK}: \texttt{gaussian}, \texttt{unityGaussian}, and \texttt{mexicanHat}. These define the amount in which a neuron $n_2$ should be updated based on an image matching neuron $n_1$. For example, the Gaussian distribution function has the form
\begin{equation}
    f(n_1, n_2) = \frac{1}{2\pi\sigma^2}\exp{\left[\frac{-(n_1-n_2)^2}{2\sigma^2}\right]},
\end{equation}
where $\sigma$ is a parameter that specifies how large a neighbourhood should be updated in each iteration. Neurons closest to the best-matching neuron are updated so they better match the training object. Neurons farther away receive less of an update. 

The damping factor, or learning rate, controls the magnitude of the update. A large damping factor leads to rapid changes in the neuron morphologies, while a low damping factor allows individual neurons to be optimized. 

The optimal training algorithms often involve multiple cycles of training. The first cycle uses the broadest distribution function and a large learning rate in order to establish rough neighbourhoods in the SOM. Subsequent cycles use either a smaller $\sigma$, a smaller damping factor, or both. An example training scheme is shown in Table \ref{tab:training}.

Several other parameters are worth addressing briefly. First, in the first cycle of training the neurons of the SOM can be initialized from random noise, or random noise with a preferred direction, instead of zeros. Next, the number of rotations can be reduced from 360 in order to speed up training time in the early cycles of training. Finally, the euclidean distance shape can be used to impose a circular mask when computing the similarity between neurons. For quadratic shapes, bright sources may enter into the masked region as the image is rotated, resulting in large variances in the euclidean distance that are unrelated to the central structure.

\begin{table}[htbp]
    \centering
    \resizebox{\textwidth}{!}{%
    \begin{tabular}{lp{0.65\textwidth}}
    \hline
    Parameter & Description \\
    \hline \hline
    \texttt{--cuda-off} & Switch off CUDA acceleration. \\
    \texttt{--dist-func}, \texttt{-f} \texttt{<string>} & Distribution function for SOM update (see below). \\
    \texttt{--euclidean-distance-dimension}, \texttt{-e} \texttt{<int>} & Dimension for Euclidean distance calculation (default = image-dimension * sqrt(2) / 2). \\
    \texttt{--euclidean-distance-type} & Data type for Euclidean distance calculation (unit8 = default, uint16, float). \\
    \texttt{--euclidean-distance-shape} & Shape of Euclidean distance region (quadratic = default, circular). \\
    \texttt{--flip-off} & Switch off usage of mirrored images. \\
    \texttt{--help}, \texttt{-h} & Print the PINK help string. \\
    \texttt{--init}, \texttt{-x} \texttt{<string>} & Type of SOM initialization (zero = default, random, random\_with\_preferred\_direction, file\_init). \\
    \texttt{--input-shuffle-off} & Switch off random shuffle of data input (only for training). \\
    \texttt{--interpolation} \texttt{<string>} & Type of image interpolation for rotations (nearest\_neighbor, bilinear = default). \\
    \texttt{--inter-store} \texttt{<string>} & Store intermediate SOM results at every progress step (off = default, overwrite, keep). \\
    \texttt{--layout}, \texttt{-l} \texttt{<string>} & Layout of SOM (cartesian = default, hexagonal). \\
    \texttt{--max-update-distance} \texttt{<float>} & Maximum distance for SOM update (default = off). \\
    \texttt{--neuron-dimension}, \texttt{-d} \texttt{<int>} & Dimension for quadratic SOM neurons (default = 2 * image-dimension / sqrt(2)). \\
    \texttt{--numrot}, \texttt{-n} \texttt{<int>} & Number of rotations (1 or a multiple of 4, default = 360). \\
    \texttt{--numthreads}, \texttt{-t} \texttt{<int>} & Number of CPU threads (default = auto). \\
    \texttt{--num-iter} \texttt{<int>} & Number of iterations (default = 1). \\
    \texttt{--pbc} & Use periodic boundary conditions for SOM. \\
    \texttt{--progress}, \texttt{-p} \texttt{<int>} & Maximal number of progress information prints (default = 10). \\
    \texttt{--seed}, \texttt{-s} \texttt{<unsigned int>} & Seed for random number generator (default = 1234). \\
    \texttt{--store-rot-flip} \texttt{<string>} & Store the rotation and flip information of the best match of mapping. \\
    \texttt{--som-width} \texttt{<int>} & Width dimension of SOM (default = 10). \\
    \texttt{--som-height} \texttt{<int>} & Height dimension of SOM (default = 10). \\
    \texttt{--som-depth} \texttt{<int>} & Depth dimension of SOM (default = 1). \\
    \texttt{--verbose} & Print more output. \\
    \texttt{--version}, \texttt{-v} & Print version number. \\
    \hline
    Distribution Functions & \\
    \hline
    Gaussian sigma damping-factor & A Gaussian distribution function. \\
    UnityGaussian sigma damping-factor & A Gaussian distribution function that is normalized such that its peak is one at the best-matching neuron. \\
    MexicanHat sigma damping-factor & A distribution function using the Mexican hat function. \\
    \hline
    \end{tabular}%
    }
    \caption{Optional flags for PINK.}
    \label{tab:pink}
\end{table}

\begin{table}[]
    \centering
    \begin{tabular}{ccccc}
        \hline
        Training & Sigma & Learning & Rotations & Iterations \\
        Stage    &       & Rate     &           &    \\
        \hline
        1 &  2.5 & 0.1 & 180 & 5 \\
        2 &  1.5 & 0.05 & 180 & 5 \\
        3 &  0.7 & 0.05 & 360 & 10 \\
        \hline
    \end{tabular}
    \caption{Parameters used for training the $10\times 10$ SOM.}
    \label{tab:training}
\end{table}

\subsection{Optimizing the training process}

Optimizing the SOM can be a difficult and uncertain process. It is difficult to ascertain whether enough training cycles have been conducted or if the SOM needs to be expanded. In the future hierarchical training schemes, which expand the SOM automatically to optimize its dimensionality, may be implemented. In the meantime, however, the user is left to rely on some useful heuristics when training their SOM. We outline some of these here.

Training a SOM involves minimizing the Euclidean distance between the SOM and the ensemble of training images. This progress can be tracked by outputting the SOM at various intervals in each training cycle (using \verb|--progress|) and mapping the training image binary onto each SOM. The total Euclidean distance is then measured relative to the final SOM to quantify its change with each step. The training cycle has converged when it no longer changes between progress steps. An example of this can be found in Fig. 3 of Galvin et al. (2019).

Visual inspection of the SOM is a useful way of assessing its status, though it is difficult to determine when the optimization process has been completed. The neurons should resemble blurred versions of the morphologies one wishes to capture. The blurring is a result of many images being averaged together. Looking at the neurons in each channel can inform, for example, whether the SOM is recognizing patterns that are irrelevant to the science in question, such as the separations between background galaxies. Addressing these issues may involve increasing the SOM's size or modifying the preprocessing step by adjusting the channel weights or masking unrelated structures.

The median and standard deviation of all Euclidean distances amongst images matched to a given neuron provides a method of inspecting whether the SOM is sufficiently large. Unless periodic boundary conditions are used, the dispersion amongst Euclidean distances will naturally increase along the SOM's edges. This is because of the coherence in a SOM. Nearby neighbourhoods are the most similar. The most complex morphologies, therefore, are often relegated to the SOM's edges so that they are connected to few other neighbourhoods. Increasing the size of the SOM should decrease the relative increase in dispersion between the edge and central neurons. 

{  Note that this is not the same as a clustering algorithm. However, the SOM could be considered a clustering approach if the distance function became extremely narrow, in which case the individual neurons would turn into independent entities.  
}

Access to a labelled test sample opens a powerful avenue to validating the SOM. Consider the science goal of automatically distinguishing between a distinct set of objects based on their morphology. With a labelled test sample, each image can be mapped onto the SOM and their label associated with their best-matching neuron. Neurons with comparable numbers of matches to multiple morphologies are ambiguous. Some ambiguous neurons are expected in regions between well-defined neighbourhoods, but too many ambiguities indicates that the SOM has not identified the morphological pattern that distinguishes those classifications.

\section{Working with the SOM files}
\label{sec:mapping}

In this Section we outline the additional functionalities that are provided by the \textsc{pyink} binary wrapper classes (\verb|pyink.SOM|, \verb|pyink.Mapping|, and \verb|pyink.Transform|). These classes can be initialized by simply providing the name of the corresponding binary file (e.g. \verb|som = pyink.SOM(som_file)|).

\subsection{SOM}

The SOM binary file contains the images for each neuron. Once it has been loaded using \verb|pyink.SOM| (for this discussion we will use the variable name \verb|som|), its data and an assortment of additional functions and properties can be accessed. The properties, which describe basic information pertaining to the shape and dimensions of the SOM and its neurons, are listed in Table \ref{tab:som}.

The data for a specific neuron can be accessed via \verb|som[| \verb|neuron]|, where \verb|neuron| is a list, array, or tuple containing the coordinates of the desired neuron. This returns an array with dimensions (neuron depth, neuron height, neuron width).
Alternatively, the entire SOM can be accessed using \verb|som.data|. This is an array in which the individual pixels from all neurons have been combined into a single image for each channel. Its dimensions are (number of channels, SOM height$\times$vertical neuron shape, SOM width$\times$horizontal neuron shape). When plotting the entire SOM it is simplest to use this array.

\begin{table}[]
    \centering
    \begin{tabular}{p{0.20\textwidth}p{0.65\textwidth}}
    \hline
    Property & Description \\
    \hline \hline
    \verb|som_rank|  &  The number of SOM dimensions. Either 2 or 3.  \\
    \verb|som_shape|  &  The size of each dimension (width, height, depth) on the SOM.  \\
    \verb|neuron_rank|  &  The number of dimensions for each neuron. Either 2 or 3. \\
    \verb|neuron_shape|  &  The size of each dimension (depth, height, width) for a neuron.  \\
    \verb|neuron_size|  &  The total number of pixels across all dimensions for a single neuron (width$\times$height$\times$depth).  \\
    \hline
    \end{tabular}
    \caption{The properties available in the \textsc{pyink} wrapper for SOM binaries.}
    \label{tab:som}
\end{table}

Two functions have been provided to plot all channels for individual neurons: \verb|plot_neuron| and \verb|explore|. The former plots only a single neuron, while the latter allows the user to use the arrow keys to move through the SOM.

{\footnotesize
\begin{verbatim}
    SOM.plot_neuron(  # Plot all channels for a single
                      # neuron.
        neuron=(0,0), # A list, array, or tuple specifying
                      # the neuron index.
        fig=None,     # A matplotlib.Figure instance to 
                      # plot the neuron.
                      # If fig=None, a new Figure will 
                      # be created.
        trim_to_img_shape=False, # Trim the neuron to 
                                 # match the dimensions 
                                 # of the images used to 
                                 # train the SOM.
        show_ticks=False  # Display the pixel indices 
                          # along the x and y axes.
    )
\end{verbatim}}

{\footnotesize
\begin{verbatim}
    SOM.explore(  # Interactively explore the SOM one 
                  # neuron at a time.
        start=(0,0), # The indices of the starting neuron.
        **kwarg   # Keyword arguments to be passed to 
                  # ``plot_neuron''.
    )
\end{verbatim}}

\subsection{Mapping}

The mapping file encodes the similarity measure between each image and each neuron. See Figure 
\ref{fig:mapping} for examples of the mapping of a VLASS source to a neuron. 
Once it has been loaded using \verb|pyink.Mapping| (here we assume the variable name \verb|mapping|), its data can be accessed through \verb|mapping.data|. This is an array with dimensions ($N$, $M_y$, $M_x$, $C$), where $N$ is the number of images and the SOM is $M_y\times M_x$ with $C$ channels. If $C=1$, the last axis is excluded. The basic properties that can be accessed using this wrapper class are provided in Table \ref{tab:mapping}.

\begin{table}[]
    \centering
    \begin{tabular}{p{0.15\textwidth}p{0.65\textwidth}}
    \hline
    Property & Description \\
    \hline \hline
    \verb|som_rank|  &  The number of SOM dimensions.  \\
    \verb|som_shape|  &  The size of each dimension (width, height, depth) on the SOM.  \\
    \verb|srcrange|  &  An array of indices corresponding to the source image axis.  \\
    \hline
    \end{tabular}
    \caption{The properties available in the \textsc{pyink} wrapper for Mapping binaries.}
    \label{tab:mapping}
\end{table}

The best-matching neuron can be accessed using \verb|mapping|\verb|.bmu|, which identifies for each image the neuron with the lowest Euclidean distance. A selection of specific indices can be supplied to obtain the best-matching neurons for only that sample. This function can also be used to identify the neuron with the Nth lowest distance, where the best-matching neuron corresponds to $N=0$. By default the function returns an array of neuron coordinates. Setting \verb|return_tuples=True| will instead return a list of tuples, which may be helpful in indexing the various \textsc{numpy} arrays.

{\footnotesize
\begin{verbatim}
    Mapping.bmu( # Return the BMU indices for each source.
        idx=None,            # The image indices of 
                             # interest. Default: all.
        N=0,                 # The Nth best-matching 
                             # neuron. Default: 0 (BMU).
        squeeze=True,        # Remove empty axes from the 
                             # returned array.
        return_idx=False,    # Include the source 
                             # index/indices as part of 
                             # the returned structure.
        return_tuples=False, # Return as a list of tuples.
    )
    Returns:
        np.ndarray -- Indices to the BMU on the SOM 
                      lattice of each source image.
\end{verbatim}}

The number of matches to each neuron can be obtained using \verb|mapping.bmu_counts|. The function returns an array of the same shape as the SOM, where each element records the number of matches for that neuron.

{\footnotesize
\begin{verbatim}
    Mapping.bmu_counts( # Return the number of matches to 
                        # each neuron.
        **kwargs # Keywords to pass to pyink.Mapping.bmu().
    )
    Returns:
        np.ndarray -- Contains the integer counts of matches
                      to each neuron. Same shape as the SOM.
\end{verbatim}}

The Euclidean distance to either the best-matching neuron or the Nth best neuron can be obtained using \verb|mapping.bmu_ed|. This function returns a 1-dimensional array with all relevant Euclidean distances.

{\footnotesize
\begin{verbatim}
    Mapping.bmu_ed( # Returns the similarity measure of 
                    # the BMU for each source.
        idx=None,   # Indices of the images to pull 
                    # information from. Default: all.
        N=0,        # Extract the Nth lowest similarity. 
                    # Default: BMU.
    )
    Returns:
        np.ndarray -- The similarity measure statistic of 
                      each image to its BMU.
\end{verbatim}}

In order to identify the sample of images that best match a specific neuron, use \verb|mapping.images_with_bmu|. The neuron indices must be supplied; it can be a tuple, list, or array. This can be used in conjunction with the $N$ keyword argument from \verb|mapping.bmu| to identify the images whose Nth closest neuron is the neuron of interest.

{\footnotesize
\begin{verbatim}
    Mapping.images_with_bmu( # Return the indices of 
                             # images that a specific BMU.
        key,      # Indices of the neuron to search for.
        **kwargs  # Additional keywords to pass to 
                  # pyink.Mapping.bmu
    )
    Returns:
        np.ndarray -- Source indices that have `key' as 
                      their BMU
\end{verbatim}}

Provided the user has a sample of labels (classifications) for their dataset, the \verb|pyink.Mapping.map_labels| function will count the number of occurrences for each label in each neuron. The return value is a dictionary with keys that are the unique labels in the provided array. Each associated item is an array with the same shape as the SOM where each element contains the number of catalog entries with that label that have been matched to the corresponding neuron.

{\footnotesize
\begin{verbatim}
    Mapping.map_labels( # Given a set of `labels', count 
                        # the number of each label matched
                        # to each neuron.
        labels,    # A 1D array containing the labels for 
                   # each item.
        idx=None,  # Indices of the images to pull 
                   # information from.
    )
    Returns:
        Dict -- Keyed by the unique `labels'. Each item 
                is an np.ndarray with the same 
                information as in Mapping.bmu_counts.
\end{verbatim}}

When training a SOM one may wish to consider its coherence. This is the count of the number of entries in which the neuron with the second lowest Euclidean distance is adjacent to the best-matching neuron (including diagonally). A higher coherence means that the neighbourhoods within the SOM are better established, which is a desired property of the SOM. {  This allows various morphological characterizations of sources, e.g., the neighbourhood of neurons that correspond to “compact doubles”, such as seen in Fig. 6. }
The coherence can be easily measured using the \verb|pyink.Mapping.coherence| function. By default this is measured from the entire sample, but a subset of indices can also be supplied. Cyclic SOMs are not supported at this time.

{\footnotesize
\begin{verbatim}
    Mapping.coherence( # Count the number of entries whose
                       # 2nd best neuron is adjacent to 
                       # the BMU.
        idx=None,      # Indices of the images to pull 
                       # information from.
        cyclic=False,  # Specifies whether the SOM is 
                       # cyclic. Not currently supported.
    )
    Returns:
        int -- Number of entries whose 2nd best neuron is 
               adjacent to the BMU.
\end{verbatim}}

The SOM is a useful tool for identifying rare populations in a sample. This is accomplished by selecting the entries with the highest Euclidean distances, i.e. the worst matches to the SOM. The \verb|pyink.Mapping.worst_matches| function selects the $N$ worst matches in either an entire sample (default) or from amongst the entries matched to a single neuron. Instead of specifying an integer value for $N$, the user may specify the fraction of the sample they wish to obtain using \verb|frac|. If this is also not specified, the entire sample will be returned. An array of the indices for the $N$ worst-matching entries is returned. If \verb|return_ed| is set to \verb|True|, an array of the corresponding Euclidean distance is also returned.

{\footnotesize
\begin{verbatim}
    Mapping.worst_matches( # Identify outlying entries 
                           # from one or all neurons.
        N=None, # Total number of indices to return. 
                # Takes priority over `frac'. If neither
                # `N' nor `frac' are provided, `N' is the
                # entire dataset.
        frac=None,   # The fraction of rows to be 
                     # returned.
        neuron=None, # Tuple of the neuron indices to 
                     # restrict the matches to. `None' 
                     # corresponds to the entire SOM.
        return_ed=False, # Return the Euclidean distance 
                         # along with the indices.
    )
    Returns:
        np.ndarray -- Indices for the worst-matching 
                      entries.
        np.nparray (optional) -- Euclidean distances for 
                                 the corresponding index.
\end{verbatim}}

\subsubsection{BMU Mask}

The \verb|pyink.Mapping| class also includes a \verb|bmu_mask| keyword argument. This can be used to exclude select neurons from all mapping computations. For example, if a set of neurons in the SOM has identified features that are not relevant, and the user wishes to exclude those instead of further optimizing the SOM, then the \verb|bmu_mask| masks out those neurons when determining the best-matching neuron.

Concretely, the \verb|bmu_mask| is a Boolean array of the same shape as the SOM. A value of \verb|True| indicates that the neuron is acceptable, with \verb|False| indicating that it should be masked out. By default all neurons are included.

The \verb|bmu_mask| can be set either when creating the \verb|pyink|\verb|.Mapping| object or afterward by setting \verb|mapping.bmu_mask| \verb|= your_bmu_mask|, \mbox{} where \verb|your_bmu_mask| is the custom Boolean array of the same shape as the SOM. If the supplied argument is a string, it will be loaded into an array using \verb|numpy.load|.

Once it has been initialized, the \verb|bmu_mask| is applied automatically any time a relevant \verb|pyink.Mapping| function is called.

\subsection{Transform}

The transform binary file contains the flip and rotation angle required to match an image to each neuron. When loaded into the \verb|pyink.Transform| class (assuming the variable name \verb|transform|), \mbox{} this information \mbox{} can \mbox{} be accessed \mbox{} through \verb|transform.data|. This is a 3-D array of axes/dimensions ($N$, $M_y$, $M_x$), where $N$ is the number of images and $M_y$ and $M_x$ are the dimensions of the SOM. Each element of the \verb|transform.data| array is a named tuple with keys ``flip'' and ``angle'', which can be accessed either by name or through standard list indexing.

The \verb|pyink.Transform| class also contains properties of the SOM, as shown in Table \ref{tab:transform}.
\begin{figure}
    \centering
    \includegraphics[width=0.99\textwidth]{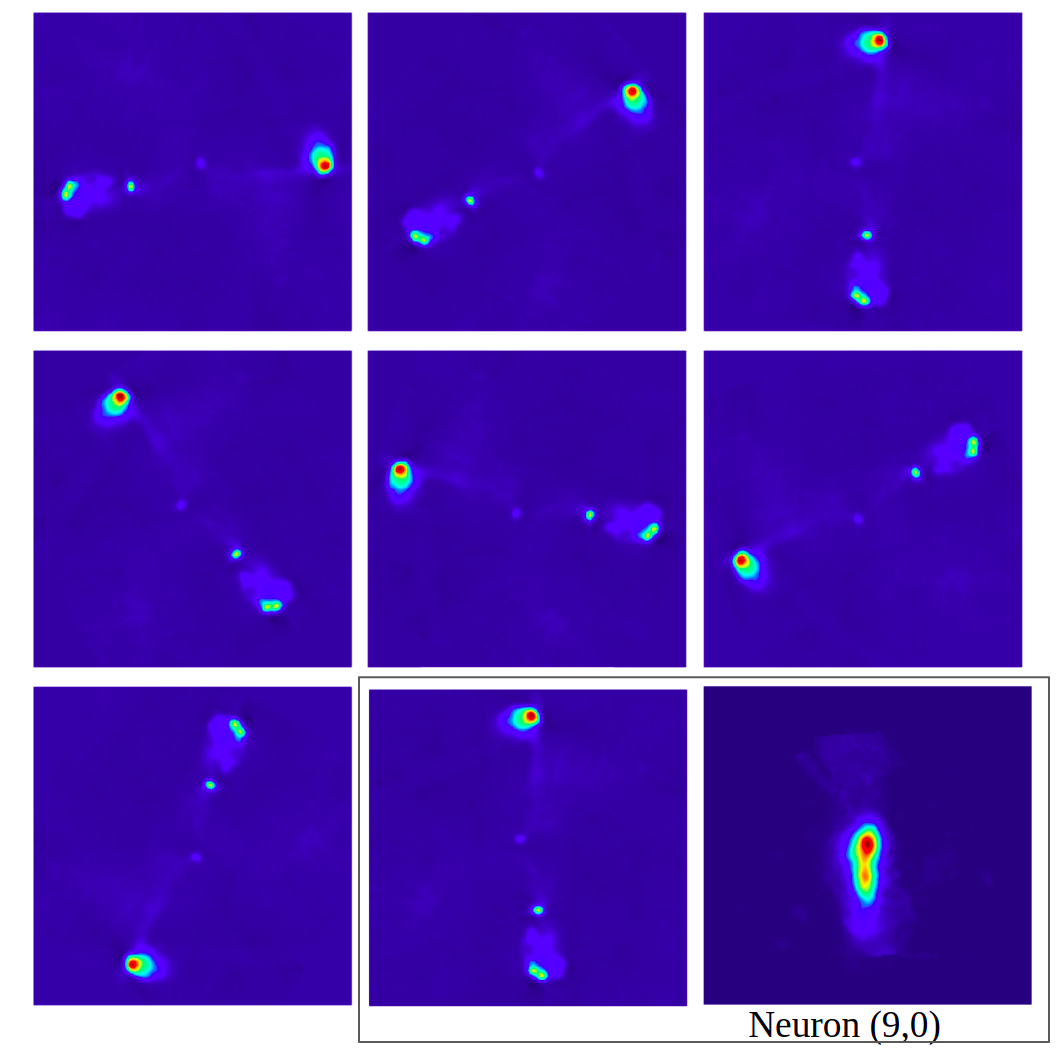}
    \caption{Example of rotational transformations achieved using the pyink modules, ``{\tt{pink\_spatial\_transform}}''. The middle bottom panel shows the transformation that best matched with the BMU i.e., the neuron (9,0) shown in the right bottom panel.}
    \label{fig:transform}
\end{figure}
\begin{table}[]
    \centering
    \begin{tabular}{p{0.15\textwidth}p{0.65\textwidth}}
    \hline
    Property & Description \\
    \hline \hline
    \verb|som_rank|  &  The number of SOM dimensions.  \\
    \verb|som_shape|  &  The size of each dimension (width, height, depth) on the SOM.  \\
    \hline
    \end{tabular}
    \caption{The properties available in the \textsc{pyink} wrapper for Transform binaries.}
    \label{tab:transform}
\end{table}

The primary use for the Transform object is in combination with the \verb|pyink.pink_spatial_transform| function, as discussed in Section \ref{sec:utilities}.

\subsection{SOMSet}

The SOMSet (\verb|pyink.SOMSet|) is a convenience class that wraps the SOM, Mapping, and Transform binaries. It is initialized via
{\footnotesize
\begin{verbatim}
    somset = pyink.SOMSet(som_file,map_file,transform_file)
\end{verbatim}}
where either the binary file names or their initialized versions can be supplied as arguments.

\subsection{SOMWriter}

The \verb|pyink.SOMWriter| class enables a user to create a SOM using custom neurons. It should first be initialized using the expected dimensions of both the SOM and its neurons.

{\footnotesize
\begin{verbatim}
    SOMWriter(  # Create a SOM using custom neurons
        binary_path,  # The output name for the SOM binary 
                      # file.
        som_shape,    # The dimensions (height, width) of 
                      # the SOM.
        neuron_shape, # The shape of each neuron (height, 
                      # width).
        comment=None, # A header message for the binary.
        clobber=False # Overwrite an existing file.
    )
\end{verbatim}}

Neurons should then be added one at a time using the \verb|SOMWriter.add| function, which accepts a single array and appends it to the binary file. The neurons are added across the columns first, then down the rows (i.e. left-to-right, top-to-bottom). For example, for a $2\times 2$ SOM, the neurons are added in the order (0, 0), (0, 1), (1, 0), (1, 1).

{\footnotesize
\begin{verbatim}
    SOMWriter.add(  # Add a neuron to the SOM binary
        img    # The neuron array to be added.
    )
\end{verbatim}}

The file should then be closed using \verb|SOMWriter.close| and can subsequently be loaded as per a normal \verb|pyink.SOM| object.

The intention of this class is to enable a user to search for images that match a certain morphology. Given an interesting image, one can preprocess the image, use it to create a new SOM, map a sample onto this SOM, and use the Euclidean distance information to identify objects that most closely match the supplied image. This is not a robust approach, but it may help limit the sample of objects to be inspected visually.

As a final note, PINK requires a width of at least 2 for the SOM. If only one image is desired, the second neuron can be either a duplicate of the first or filled with zeroes.

\subsection{Image Visualization}

Plotting a single image from the image binary can be done using \newline \verb|pyink.plot_image|. The only required input is the image binary, in which case one of the images is chosen at random. The user may supply an integer index (\verb|idx|) instead to choose which image to plot. All image channels are plotted in a series of horizontal panels in a single figure using \verb|matplotlib| colormaps specified by \verb|cmaps|.

If a \verb|pyink.SOMSet| is specified then the best-matching neuron can be plotted below the image by setting \verb|show_bmu=True|. It is recommended to set \verb|apply_transform=True| so that the image is transformed to match the neuron.

{\footnotesize
\begin{verbatim}
plot_image(   # Plot an image from the image binary.
    imbin,    # Image binary
    idx=None, # (int) Index of the image to plot. If None, 
              # one is chosen at random.
    df=None,  # pandas.DataFrame with information on the 
              # sample. If set and idx is None, an index 
              # will be chosen from the DataFrame. Will 
              # attempt to name the plot using the 
              # "Component_name" column.
    somset=None, # (SOMSet) Container holding the SOM, 
                 # mapping, and transform.
    apply_transform=False, # Transform the image to 
                           # match the neuron.
    fig=None, # Existing pyplot.Figure for plotting. 
              # If None, one is created.
    show_bmu=False, # Display the best-matching neuron 
                    # alongside the image.
    wcs=None, # An astropy.wcs.WCS axis to apply to the 
              # image.
    grid=False, # Show grid lines on the plot.
    cmaps,    # A list of pyplot colormaps to iterate 
              # through. If the list is shorter than 
              # the number of channels, it will be 
              # cycled through. Default: ["viridis", 
              # "plasma", "inferno", "cividis"]
)
\end{verbatim}}

\subsection{Additional Utilities}
\label{sec:utilities}

A number of utility functions have been included in \textsc{pyink}. 
The \newline \verb|pyink.pink_spatial_transform| function transforms an image based on its flip and angle from the Transform binary to match it to the neuron frame.
See Figure \ref{fig:transform} for an example of rotational transformations. 
Alternatively, the inverse transform can be applied to a neuron to match it to an image. This is sometimes beneficial as the neurons are larger than the image cutouts because rotating a square image requires a larger footprint. The rotated neuron can then easily be trimmed to match the size of the image. 

{\footnotesize
\begin{verbatim}
    pyink.pink_spatial_transform( # Apply the PINK spatial 
                                  # transformation to an 
                                  # image
        img,        # Image to spatially transform
        transform,  # Spatial transformation specification 
                    # following the PINK standard
        reverse=False # Apply the transform in the 
                      # opposite order, for example to
                      # match a neuron to an image
    )
    Returns:
        np.ndarray -- Spatially transformed image
\end{verbatim}}

The following utilities are all used in conjunction with image-based annotations, which are discussed in Section~\ref{sec:image-antns}. Given an image mask of an arbitrary shape, \verb|pyink.compute| \verb|_distances_between_valid_pixels| considers the valid pixels in a pair-wise fashion, computing the distance between each pair. See \verb|scipy.spatial.distance.cdist| for more on how this is performed. For a mask containing $P$ valid pixels, the result is a $P\times P$ array.

{\footnotesize
\begin{verbatim}
    pyink.compute_distances_between_valid_pixels( # Given a 
        # mask, compute the distance between each pixel in
        # a pair-wise fashion
        mask # 2D Boolean array to compute distances 
             # between
    )
    Returns: numpy.ndarray -- Matrix object of distances 
                              between pixels (see
                              scipy.spatial.distance.cdist)
\end{verbatim}}

A more useful representation of  these distances can be obtained from \verb|pyink.distances_between_valid_pixels|. This \mbox{\;\;} uses \mbox{\;\;} \newline \verb|pyink.compute_distances_between_valid| \verb|_pixels| to compute the pair-wise distance matrix and then determines the maximum separation between any two pixels, the indices of those pixels, and also provides the distance matrix itself. This provides a measure of the maximum extent of a source, provided the mask encodes the location of flux within the image.

{\footnotesize
\begin{verbatim}
    pyink.distances_between_valid_pixels( # Given a mask, 
        # compute the distances between all valid pixels 
        # and return the maximum separation between any 
        # two pixels, which pixels these were, the actual 
        # distances between each pair-wise combination 
        # of valid pixels.
        mask # 2D Boolean array to compute distances 
             # between
    )
    Returns: Tuple(float, np.ndarray, np.ndarray) -- 
        Maximum separation between any two pixels, indices 
        of those pixels, and distance matrix
\end{verbatim}}

Given a \verb|pyink.Filter| object, which is used in image-based annotations, the \verb|pyink.valid_region| function constructs a mask based on the pixels that have been assigned certain labels. This function can also exclude certain labels from the mask.

{\footnotesize
\begin{verbatim}
    pyink.valid_region( # Constructs a valid region based 
                        # on whether labels are present 
                        # or not.
        filter,          # Filter to use as the base
        filter_includes, # Labels to include in the 
                         # masked region
        filter_excludes  # Labels to exclude from the 
                         # masked region
    )
    Returns:
        np.ndarray -- Boolean masked constructed following 
                      the specifications
\end{verbatim}}

Finally, the \verb|pyink.area_ratio| function can be used to estimate the relative sizes between two masks. Given a pair of filters, it creates a mask for each one, counts the number of valid pixels, and then takes the ratio between those values. For a concrete example of where this is useful, consider a SOM trained on 2-channel (radio and IR) data. If the area ratio (radio/IR) is large, then the radio source is large while the position of the IR host is well-determined. This can then be used to judge which neurons are ``best''. 

{\footnotesize
\begin{verbatim}
    pyink.area_ratio( # Compute the order of the filters 
                      # by the relative ratio between 
                      # desired regions in each filter
        filter1, # The first filter
        filter2, # The second filter
        filter_includes=None, # Labels to include in the 
                              # masked region
        filter_excludes=None, # Labels to exclude from the 
                              # masked region
        empty_check=True, # Return 1 if `filter2` has no 
                          # valid pixels
    )
    Returns:
        float -- The ratio between the valid areas
\end{verbatim}}

\subsubsection{Sampling Neurons}

The \verb|pyink.SOMSampler| class provides a way to select a subset of the neurons from the SOM. The best application for this tool is to enable the visualization of a subset of the SOM's neurons. The object can be initialized by specifying the SOM and the number of neurons to be sampled. The sampled neurons are then contained in \verb|SOMSampler.points|, which is a list of neuron indices.

{\footnotesize
\begin{verbatim}
pyink.SOMSampler( # Creates a list of sampled neurons.
    som, # The SOM from which neurons are chosen
    N,   # The number of neurons to sample
    method="kmeans", # The sampling function. Either 
                     # "kmeans" or "random".
    **kwargs, # Keyword arguments to be passed to 
              # the sampling function.
)
\end{verbatim}}

The two methods that are currently available for sampling neurons are ``random'' and ``kmeans'' (default). The random sampler {  is provided with a target number of sampled neurons,} \verb|N|, and populates the list of sampled neurons one at a time. A neuron is chosen at random and added to the list only if it does not lie within \verb|min_dist| of another sampled neuron.  If this procedure fails at least \verb|max_attempts| times, no more neurons will be sampled. This can result in { a final list} of fewer than \verb|N| sampled neurons.

{\footnotesize
\begin{verbatim}
SOMSampler.random_sampler( # Randomly sample the neurons.
    min_dist=2,      # Minimum separation between each 
                     # sampled neuron.
    max_attempts=10  # Maximum number of attempts at 
                     # adding a new neuron. If this 
                     # number is  reached, no more sampled 
                     # neurons will be added. This may 
                     # result in fewer than `N' sampled 
                     # neurons.
)
\end{verbatim}}

The k-means sampler employs k-means clustering \citep{lloyd1982} to partition the SOM into \verb|N| regions. This provides a much more uniform spread of neurons than the random sampling, so is the recommended method. A \verb|pyink.Mapping| object can be supplied in order to weight each cluster by the number of matches to each neuron. This will increase the number of sampled neurons in SOM neighbourhoods with a large number of matches.

{\footnotesize
\begin{verbatim}
SOMSampler.kmeans_sampler( # Use k-means clustering to 
    # optimize the position of N points on the SOM.
    mapping=None, # (pyink.Mapping) Weight the clusters 
                  # by the frequency of each neuron.
)
\end{verbatim}}

\section{Annotations}
\label{sec:annotations}

The analyses that are possible following the training of a SOM all incorporate an annotation process where the user attaches an ``annotation'' (a ``label'' or ``classification'') to each neuron. This utilizes the dimensionality reduction provided by the SOM. Manually classifying each input image, or even a statistically useful subset, is a major undertaking that is rarely feasible. The user can instead annotate each neuron in the SOM. Then, when a preprocessed image is matched to a neuron, its corresponding catalog entry inherits all annotations that were attached to the neuron. The user need only annotate hundreds of neurons instead of thousands of images. These annotations can take a multitude of forms, including one or more manually-defined text labels per neuron, text labels attached using knowledge transfer from a labelled dataset, or 2-dimensional masks that are associated with their own label.


\subsection{Manual text labels}

The simplest annotation is a collection of text labels applied to each neuron. For example, the user may have six morphological classifications for radio galaxies. Each neuron, and by extension the radio components that are matched to it, is nominally assigned one of these labels.

Creating these labels is best done manually using a text editor or spreadsheet. Each row can be uniquely specified by the neuron indices, and the user is free to supply whichever annotations they wish. This can subsequently be joined with the original catalog after it has been updated with a column for the best-matching neuron. The visualization tool \verb|pyink.SOM.explore| is useful for navigating through the SOM while annotating the neurons in a separate file.

\subsection{Knowledge transfer from a labelled dataset}

In some circumstances a labelled dataset already exists and can be used to define each neuron's annotation(s). \citet{Mostert2021}, for example, used labelled LoTSS data to determine the prevalent classes for each neuron in their SOM. Provided the labelled dataset is representative, the most common label within a neuron can be adopted as its annotation. If multiple labels have a comparable number of occurrences, either all may be adopted or the neuron can be flagged as ambiguous.

The process to perform this knowledge transfer follows the standard steps. Image cutouts for each entry in the labelled dataset should first be preprocessed and then mapped onto the SOM. The best-matching neuron should then be identified. For a given neuron, count the occurrences of each label in the dataset. This can be used to compute the fraction of each label assigned to each neuron. One should ensure that the total number of entries for each label is considered before adopting the annotations (i.e. a $50/50$ split is ambiguous).

The function \verb|pyink.Mapping.map_labels| is provided to simplify the knowledge transfer. It performs the occurrence counting described above. Given a list of labels, the function returns a dictionary in which each label (the dict key) is associated with an array counting the number of times that label has been assigned to the corresponding neuron. The statistics and choices of annotations are left to the user.

\subsection{Image-Based Annotations}
\label{sec:image-antns}

A more complex annotation process involves the creation of any number of 2-dimensional masks (also known as ``filters''). Each is assigned one or more labels. As \textsc{PINK} records the transformations required to match a preprocessed image to each neuron, the inverse transform can be applied to the best-matching neuron and all of its masks in order to convert them to the frame of the input image. The labels can be interpreted in a number of different ways during the collation process (see Section \ref{sec:collation}). This can, for example, be used to group together radio components belonging to the same astronomical source.

\subsubsection{The annotation interface}
\label{sec:antn}

An interface designed to enable the creation of labelled masks is provided \mbox{} by \mbox{} \textsc{pyink} \mbox{} in ``Example\_Scripts/run\_annotation \_example.py''. It can be launched from the command line as follows.
%
%
{\footnotesize
\begin{verbatim}
run_annotation_example.py [-h] [-d] [-k KEY [KEY ...]] [-r
[RESULTS]] [-c] som

positional arguments:
  som                   Path to the desired SOM to annotate

optional arguments:
  -h, --help            show this help message and exit
  -d, --dont-save       The default behaviour of the 
                        `Annotator` class is to save when 
                        moving between neurons (by default 
                        appending `.results.pkl` to the 
                        SOM path). This disables the
                        behaviour. (default: False)
  -k KEY [KEY ...], --key KEY [KEY ...]
                        The key of the neuron to update. 
                        Following the scheme from 
                        `Annotator` this will be converted
                        to a `tuple` with elements of type 
                        `int`. (default: None)
  -r [RESULTS], --results [RESULTS]
                        The path to a previously saved 
                        annotation set. If a path is 
                        provided attempt to load from it. 
                        If just the option flag is 
                        presented assume the desired file 
                        follows the default naming scheme 
                        (see the `--save` option). 
                        Otherwise, do not attempt to load 
                        any existing results file. 
                        (default: False)
  -c, --resume          Continue the annotation process 
                        from the first un annotated neuron, 
                        skipping those already labeled
                        (default: False)
\end{verbatim}}

\begin{figure}
    \centering
    \includegraphics[width=0.99\textwidth]{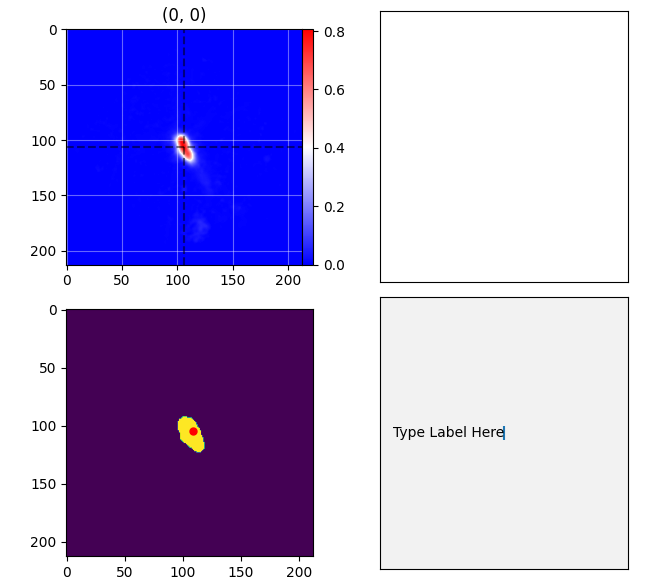}
    \caption{The graphical interface for creating masks. New labels are added in the bottom right panel. Once added, it will appear in the top right panel next to a checkbox. To create a new mask, select all applicable labels and then either left click on the signal in the top panel of the appropriate channel (when there are multiple) or right click and hold to draw a lasso region around the flux. The masked region is shown in the bottom panel. The click position is indicated by the red dot.}
    \label{fig:interface}
\end{figure}

An example of the interface is shown in Fig. \ref{fig:interface}. It consists of two ``image panels'' for each channel in the preprocessed images plus two ``label panels'' used for labels. The top image panel(s) displays the preprocessed image(s), while the bottom is used to display the mask. Each of the image panels can be panned or zoomed using the normal \verb|matplotlib| graphical interface tools. The bottom label panel is used for adding new labels, which will appear in the top panel beside a checkbox.

Masks are created using the following steps.
\begin{enumerate}
    \item Add label options using the bottom right label panel. This will populate the top label panel.
    \item Select the checkboxes for all desired labels.
    \item Create the mask. This can be done in two ways.
    \begin{enumerate}
        \item Click on a pixel within a group of bright (nonzero) pixels. This fills in all pixels connected to the click position that are at least 2 times the standard deviation of pixel intensities in the corresponding image. This threshold can be adjusted in increments of 0.5$\sigma$ using the up and down arrow keys.
        \item Right click and hold to draw a lasso around the structure of interest.
    \end{enumerate}
    \item Repeat for any other masks.
    \item Press \verb|n| to continue to the next neuron. Repeat for all neurons.
\end{enumerate}

Note that the interface will only register a key press if the cursor is inside one of the panels. The full list of key commands is listed in Table \ref{tab:antn_keys}.

\begin{table}[htb]
    \centering
    \begin{tabular}{p{0.1\textwidth}p{0.7\textwidth}}
    \hline
    Key & Description \\
    \hline \hline
    \verb|q| & quit the application  \\
	\verb|c| & clear the state of all masks, including clicked filled regions and lasso regions \\
	\verb|d| & remove the last valid click added across all masks \\
	\verb|n| & move to the next neuron for annotation \\
	\verb|b| & move to the previous neuron for annotation \\
	\verb|u| & enable or disable the live updating of the island segmentation around click positions (default is enabled) \\
	\verb|`up'| & increase the sigma-clipping threshold by 0.5 sigma (1 sigma = std. dev. of all pixel intensities, default starting value is 2-sigma) \\
	\verb|`down'| & decrease the sigma-clipping threshold by 0.5 sigma (1 sigma = std. dev. of all pixel intensities, default starting value is 2-sigma) \\
    \hline
    \end{tabular}
    \caption{The key commands available in the annotation interface. These will only be registered if the cursor is inside one of the panels.}
    \label{tab:antn_keys}
\end{table}

When a label is created it is assigned a unique prime number. For each label given to a pixel, the pixel's intensity is incremented by the corresponding prime number. This allows all masks to be stored in a single array in order to keep the output file small. 
As a result, the same pixel can be included in any number of masks without being overwritten. The \verb|matplotlib| graphical interface displays the intensity of the pixel under the cursor, allowing the labels to be inspected to ensure everything is working as intended.

Known issues:
\begin{itemize}
    \item Lasso regions (drawn with the right mouse button) cannot be undone. The entire collection of masks must be cleared by pressing the `c' key.
    \item For large images (in terms of the number of pixels) and large SOMs, the memory required to annotate the SOM may exceed the available RAM. This will crash the annotator and corrupt all progress. Quitting and relaunching the application usually decreases the memory usage and may help the user complete the annotation process. Keeping regular backups of the output file (default is ``som\_name .records.pkl'') is recommended.
    \item Overlapping masks with different labels may result in undesired behaviour. To make sure this is not happening, look at the values of the pixels in question by hovering the cursor over them. The desired pixel value is the product of all integers corresponding to the assigned labels.
\end{itemize}

\subsubsection{Collation}
\label{sec:collation}

Collation is the process of transferring neuron annotations to the input catalog. In addition to the text labels discussed in the previous sections, annotations based on masks can handle much more complex functionality. This includes, for example, grouping together entries that are projected into a common mask.
The collation process consists of the following steps.
\begin{enumerate}
    \item Create a filter to project the masks into the reference frame of the input images.
    \item Define the actions to be taken for each mask label.
    \item Sort the entries and/or neurons to prescribe which groups should be created first.
    \item Create groups of entries based on their associated masks and the actions defined for those masks.
\end{enumerate}

A ``filter'' is the annotation mask described in Section \ref{sec:antn}. Before it can be applied to an input image, the filter must first be transformed (rotated and flipped) to align with the image. This is accomplished using \texttt{pyink.FilterSet}. Given a list of the central coordinates for each preprocessed image as well as a similar list for each channel, \texttt{pyink.FilterSet} identifies and keeps track of all of the supplied coordinates (in \texttt{match\_catalogs}) that are projected into the mask(s) of their corresponding channel. To speed up the computation it only considers entries within a specified separation (\texttt{seplimit}) of the central entry.

{\footnotesize
\begin{verbatim}
FilterSet(
    base_catalog, # Coordinates for the center of each 
                 # image associated with the Mapping 
                 # and Transform objects in the SOMSet. 
                 # astropy.coordinates.SkyCoord
    match_catalogs, # List of coordinates to project 
                 # through the masks. One list per 
                 # channel.
    annotation,  # The annotated SOM
    som_set,     # A SOMSet containing the SOM, Mapping, 
                 # and Transform objects.
    cpu_cores=None, # Number of CPU cores to use during 
                 # projection. Default: 1.
    seplimit=1*u.arcminute, # Angle around each component 
                 # to consider matches,
                 # based on `search_around_sky'.
                 # Default: 1*astropy.units.arcminute
    progress=False, # Print a progress bar.
    **ct_kwargs, # Additional keywords for 
                 # pyink.CoordinateTransformer
)
\end{verbatim}}

The labels used to create the masks can be anything the user wishes. These must be assigned an ``action'' in order for the grouping stage to understand how they should be interpreted. This is done used \texttt{pyink.LabelResolve}, which takes as input a dictionary that links each label to an action. The full list of actions is provided in Table \ref{tab:actions}. 
These correspond to specific actions in a \textsc{networkx} graph. A \verb|NODE| is created for each catalog entry. An \verb|EDGE| is used to join together multiple nodes in order to create a group. This provides an efficient way of linking together multiple entries.

For the example of radio components, \verb|LINK| is used to join radio components into sources, \verb|DATA_ATTACH| attaches an IR host to the radio source, and \verb|ISOLATE| can optionally be used to prevent sidelobes from being incorporated into the source. If a morphological label is also included in the mask, it can be incorporated into the source by using \verb|TRUE_ATTACH|.

{\footnotesize
\begin{verbatim}
LabelResolve(
    dict  # The Action to be taken for each label.
)
\end{verbatim}}

The order in which items are collated has a direct impact on the output catalog. To understand the reasoning behind this, consider an example in which a SOM has been trained on radio and IR images with the intent of grouping radio components into complex sources. For an astronomical source that should contain three components, it is the central component that will best represent the overall structure of the source. Each component is mapped separately onto the SOM, and an image centered on a different component may not encompass the component farthest from it. If one were to collate these components in a random order, it may result in two sources -- one with two components and the third component as a separate source. To address this the neurons should be sorted by a metric that attempts to prioritize components at the centers of their respective source. In this example that could be by choosing the neuron with the largest structure or by considering the relative area between radio and IR neurons (i.e. prioritize large radio sources with a well-defined IR host position).

The sorting of entries and/or neurons is performed using the \texttt{pyink.Sorter} class. 

\begin{longtable}{|p{0.2\textwidth}|p{0.77\textwidth}|}
    \caption{The commands that can be used to interpret the annotations. These are defined in ``pyink.Action'', and are applied against either the NODE or EDGE of a networkX graph. The GreedyGraph performs the operation described with each command when iterating across each of the projected filters. Internally, a networkX.MultiGraph is established, with a single NODE representing a single row (component) in the base catalog. An EDGE connects two NODEs together, which signifies these are two components of a single source. Both a NODE and a EDGE have the ability to store any key-value pair under their corresponding data attribute.  }
     \label{tab:actions} \\
    \hline
    Command & Description \\
    \hline \hline
    \endfirsthead 
    
    \multicolumn{2}{c}%
    {{\bfseries Table \thetable\ continued from previous page}} \\
    \hline
    Command & Description \\
    \hline \hline
    \endhead 
    
    \hline
    \multicolumn{2}{r}{{Continued on next page}} \\ 
    \endfoot
    
    \hline
    \endlastfoot
    \verb|LINK| & Create an \verb|EDGE| between all \verb|NODE|s that pass together through a mask. \\
    \verb|UNLINK| & Destroy any \verb|EDGE|s that may be connected to any \verb|NODE|s that pass through a mask.  \\
    \verb|RESOLVE| & Sets the ``Resolve'' field to \verb|True| in the data dictionary that is attached to edges of a set of \verb|NODE|s.  \\
    \verb|FLAG| &  Sets the ``Flagged'' field to \verb|True| in the data dictionary that is attached to edges of a set of \verb|Node|s. \\
    \verb|PASS| & Do not take any actions relating to this label.  \\
    \verb|NODE_ATTACH| & Attach the value of `True' to each of the \verb|NODE|s on the graph of all sources that passed through a mask under a ``key'' corresponding to the current label value. \\
    \verb|DATA_ATTACH| & Attach the numeric index of all sources that passed through a mask to the data attribute of a (set of) \verb|EDGE| data component under a ``key'' corresponding to the current label value.     \\
    \verb|TRUE_ATTACH| & Add the value of ``True'' to a key corresponding to the current label value to the \verb|EDGE| data attribute to all \verb|EDGE|s created by the projection of this mask.    \\
    \verb|FALSE_ATTACH| & Add the value of ``False'' to a key corresponding to the current label value to the \verb|EDGE| data attribute to all \verb|EDGE|s created by the projection of this mask.   \\
    \verb|ISOLATE| & Remove all \verb|EDGE|s completely from any sources that pass through the projection of this mask.  \\
    \hline
    \end{longtable}
{\footnotesize
\begin{verbatim}
Sorter(
    som_set, # SOMSet container holding the SOM, Mapping, 
             # and Transform files.
    *args,   # Arguments to be passed to the corresponding 
             # ordering function.
    mode=`best_matching_first',  # Sorting mode operation 
             # (Options: `best_matching_first', 
             # `largest_first', and `area_ratio')
    **kwargs # Keywords to be passed to the corresponding 
             # ordering function.
)
\end{verbatim}}
It contains three built-in methods (\verb|mode|) that may be useful. Each method prioritizes entries in a different way in an attempt to produce the most accurate groups.
\begin{itemize}
    \item \verb|best_matching_first| -- Begin with the entries whose preprocessed images are the most similar to their respective neuron.
    \item \verb|largest_first| -- Begin with the largest size amongst all neurons, as determined by the maximum separation between any two non-zero pixels in a specified channel.
    \item \verb|area_ratio| -- Begin with the neurons which have the largest size ratio between two specified channels. The user can indicate which masks should be considered in the size computation for each channel.
\end{itemize}
A custom \verb|mode| can be implemented by inheriting from the class and defining the new function. We show this below by implementing a random sorting function.
%
%
%
%
{\footnotesize
\begin{verbatim}
import numpy as np
import pyink as pu

class CustomSorter(pu.Sorter):
    NEW_MODES = ["random"]

    def __init__(
        self, 
        som_set, 
        *args, 
        mode="random", 
        **kwargs
    ):
        self.MODES += self.NEW_MODES
        super().__init__(
            som_set, 
            *args, 
            mode=mode, 
            **kwargs
        )
        print(self.som_set)

        if mode == "random":
            self.order = self._random_order(
                *args, 
                **kwargs
            )

    def _random_order(self):
        order = np.arange(0, self.mapper.data.shape[0], 1)
        np.random.shuffle(order)
        return order
\end{verbatim}}

The grouping stage is the process that performs the collation. Its aim is to group together as many catalog entries as possible with as few mappings as possible. 
Grouping uses a ``greedy graph'' that is called upon the creation of a \verb|pyink.Grouper| object. This class requires, at minimum, the filters (\verb|FilterSet|), annotations (\verb|Annotator|), actions (\verb|LabelResolve|), and sorter (\verb|Sorter|) that will be used for collation. In addition, the user may supply a function to the \verb|src_stats_fn| keyword argument which takes as input the index return by the \verb|Sorter| class and returns a dictionary of information to be attached to the group. The example below shows how to attach the best-matching neuron to a group for the first entry assigned to it. Note that this function must be defined locally so that the function has access to all variables and functions that it tries to use.

{\footnotesize
\begin{verbatim}
Grouper(
    filters,       # The projected cookie-cutter filters 
                   # (pyink.FilterSet)
    annotations,   # Annotated filters (pyink.Annotator)
    label_resolve, # Actions to perform for individual 
                   # labels (pyink.LabelResolve)
    sorter, # Specifies the order which the filters are 
            # iterated over (pyink.Sorter)
    src_stats_fn=None, # User-provided function passed 
                       # to `greedy_graph`
    progress=False  # Provide a `tqdm' style progress bar
)
\end{verbatim}}

%
{\footnotesize
\begin{verbatim}
def src_fn(idx):
    # Note: som_set is defined locally, but still used 
    # here.
    return {'idx': idx,
            'bmu': som_set.mapping.bmu(idx)}

group = pu.Grouper(filters, annotation, actions, sorter, 
                   src_stats_fn=src_fn, progress=True)
\end{verbatim}}

Once the \verb|Grouper| object has been created, the desired properties for each of the resulting subgroups can be extracted and compiled into an output table. The specifics of this step depend significantly on the quantities of interest, but several aspects are the same. We provide a detailed example below. Here we are linking together all radio components that are projected into the same mask (labelled ``Related Radio'') and attaching all IR sources projected into another mask (labelled ``IR Host''). The groups are sorted by the number of radio components they contain, and the information supplied through the user-defined statistics function is extracted using \verb|group.graph.edges(data=True)| along with the index of the subgroup. We return an \verb|OrderedDict| so that the results can be easily converted into a \verb|pandas.DataFrame|.

{\footnotesize
\begin{verbatim}
import pandas as pd
import numpy as np
from collections import OrderedDict
import networkx as nx
import pyink as pu

def source_info(group, subg, base_cat, match_catalogs):
    '''The function that extracts information from a 
    networkx group. This examples links together radio 
    components and attaches a host galaxy.
    '''
    
    G = group.graph
    radio_cat, ir_cat = match_catalogs
    somset = group.sorter.som_set
    subg = list(subg) # An index for the group, based on
                      # `match_catalog'

    data = [d for d in G.edges(nbunch=subg, data=True)]
    data = min(data, key=lambda x: x[2]["count"])

    # `data' contains both `src_idx' from the Sorter 
    # class and any  user-defined quantities in 
    # `src_stats_fn' provided to the Grouper.
    src_idx = data[2]["src_idx"]
    ir_hosts = data[2]["IR Host"]
    bmu = tuple(data[2]["bmu"])
    filters = group.filters[src_idx]
    
    # Radio position info
    radio_ind_mask = filters[0].coord_label_contains(
        "Related Radio")
    radio_inds = filters[0].coords.src_idx[radio_ind_mask]
    best_comp = base_cat.iloc[src_idx]
    radio_comps = radio_cat.iloc[radio_inds] # networkx 
                                       # resets numbering
    comp_names = list(radio_comps["Component_name"])

    # IR position info
    ir_comps = ir_cat.iloc[ir_hosts]
    ir_ra = list(ir_comps["RA"])
    ir_dec = list(ir_comps["DEC"])

    # Derived info
    total_flux = radio_comps["Total_flux"].sum()
    euc_dist = somset.mapping.bmu_ed(src_idx)[0]
    
    source = OrderedDict(
        src_idx=src_idx,
        RA_source=best_comp["RA"],
        DEC_source=best_comp["DEC"],
        N_components=len(radio_comps),
        Best_component=best_comp["Component_name"],
        Component_names=comp_names,
        N_host_candidates=len(ir_comps),
        RA_host_candidates=ir_ra,
        DEC_host_candidates=ir_dec,
        Best_neuron=bmu,
        Euc_dist=euc_dist,
        Total_flux=total_flux,
    )

def collate(group, *args, **kwargs):
    # Loop through the groups, starting with the one 
    # with the most components.
    G = group.graph
    for i, subg in enumerate(
        sorted(
            nx.connected_components(G), 
            key=lambda x: len(x), 
            reverse=True
        )
    ):
        yield source_info(group, subg, *args, **kwargs)

def src_fn(idx):
    # idx is taken from the Sorter, so one per image
    return {"idx": idx,
            "bmu": somset.mapping.bmu(idx),
            "ra": radio_posns[idx].ra.deg,
            "dec": radio_posns[idx].dec.deg,
    }

group = pu.Grouper(
    filters, 
    annotation, 
    actions, sorter, 
    src_stats_fn=src_fn, 
    progress=True
)

# Create a pandas.DataFrame catalog
final_cat = pd.DataFrame(
        collate(
            group, 
            radio_cat, 
            (full_radio_cat, ir_cat), 
            **kwarg
        )
)
\end{verbatim}}

\section{Hardware Requirements}
\label{sec:hardware}

Training a SOM using PINK is computationally expensive. To address this, PINK takes advantage of the parallel processing capabilities of modern Graphical Processing Units (GPUs). Access to a GPU is not strictly necessary, but is strongly recommended.

The total memory requirement for the GPU is dictated by the total number of floating point numbers (32 bit) that need to be stored in the SOM data array and a single image with all of its transformations. 
An $N\times N$ image requires $\sqrt{2}$ times the number of pixels along each dimension to support its transformations. In total there are $2R$ transformations, where $R$ is the number of rotations and the factor of $2$ accounts for the horizontal flip. For images with $C$ channels and an $M\times M$ SOM, the total memory requirement is
\begin{equation}
    \label{eqn:mem}
    Mem = 54.5 \left( \frac{R}{360}\right) C \left( \frac{N}{100} \right)^2
    + 7.6 \left( \frac{M}{10}\right)^2 C \left( \frac{N}{100} \right)^2
    \textsc{MB}.
\end{equation}
This can approach the gigabyte regime for large images and/or SOMs. Note that matching images to a trained SOM has the same memory requirements as training the SOM, but is much faster.




\section{Case Study: Using the SOM to Identify Sidelobes in Radio Images}\label{sec:sidelobes}


Here we walk through an example project that was conducted using the SOM. The goal was to quantify the probability that a radio component in the VLASS component catalog\footnotemark\ \citep{Gordon2021} is a false positive detection originating from a sidelobe. 
The VLASS QuickLook images are the result of rapid image processing methods that allow data access in a timely fashion, but at the expense of image quality \citep{Lacy2019, Lacy2020}.
A consequence of this rapid image processing is that prominent sidelobes are often present in the VLASS QuickLook images\textemdash especially near bright sources\textemdash and these spurious detections can be picked up by source finders \citep[see also Section 2.3 of][]{Gordon2021}.
Training a SOM to identify these sidelobes thus presents an opportunity to improve the fidelity of source catalogs.
The process we adopt can be followed in the form of a Jupyter notebook tutorial in the ``pyink/Example\_Notebooks'' directory. All supplemental files -- the sample catalog, SOM, and annotations -- are provided at
\url{https://cirada.ca/vlasspipeline#pipeline3}.
\footnotetext{\url{https://cirada.ca/vlasscatalogql0}}

\subsection{Training a sidelobe SOM}
The first step in this process is to choose the training sample. 
Using the catalog described in \citet{Gordon2021}, we select only components with a \verb|Quality_flag| of either 0 or 1.
Such components have passed all the quality assurance tests described in Section 2 of \citet{Gordon2021} apart from the ``peak-to-ring'' ratio -- a metric specifically designed to identify potential sidelobes.
Explicitly, the ``peak to ring'' measures the ratio of the maximum brightness within 5" of the component to the maximum brightness in an annulus centred on the component with inner and outer radii of $5''$ and $10''$ respectively centred on the component position.
We then choose those components with ``peak-to-ring'' $\leq 3$, and from the resulting table we randomly sample 100,000 components to use for training. The object classification is thus biased towards sidelobes and is not representative of the general population of sources detected in VLASS QL1.

%
{\footnotesize
\begin{verbatim}
import pandas as pd

sample = pd.read_csv(cirada_catalog)
sample = sample[sample.Quality_flag.isin([0,1])]
sample = sample["Peak_to_ring"] < 3
sample = sample.sample(100000)  # choose 100k components 
                                # at random
sample = sample.reset_index(drop=True)  # Reset the 
                                # DataFrame indexing.
\end{verbatim}}

Image cutouts were then downloaded for each of these components using the Legacy SkyViewer\footnotemark. Each was 150$\times$150 pixels and measured $1.5'$ on each side.
\footnotetext{\url{legacysurvey.org/viewer}}
These image cutouts were preprocessed using the following steps. This creates both an image binary and a file with name ending in ``.records.pkl'', which contains a list of the catalog indices that were successfully preprocessed.
\begin{enumerate}
    \item Estimate the rms in an image cutout
    \item Mask out all values below a signal-to-noise ratio of 2
    \item Apply a log scaling to the remaining data
    \item Scale the data on a 0 to 1 scale
\end{enumerate}

{\footnotesize
\begin{verbatim}
import os
from tqdm import tqdm
import numpy as np
from astropy.io import fits
from astropy.wcs import WCS
import pyink as pu


def load_fits(filename, ext=0):
    hdulist = fits.open(filename)
    d = hdulist[ext]
    return d


def load_radio_fits(filename, ext=0):
    """Load the data from a single extension of a fits 
    file."""
    hdu = load_fits(filename, ext=ext)
    wcs = WCS(hdu.header).celestial
    hdu.data = np.squeeze(hdu.data)
    hdu.header = wcs.to_header()
    return hdu


def scale_data(data, log=False, minsnr=None):
    """Scale the data so that the SOM behaves 
    appropriately."""
    img = np.zeros_like(data)
    noise = pu.rms_estimate(
        data[data != 0], 
        mode="mad", 
        clip_rounds=2
    )
    # data - np.median(remove_zeros)

    if minsnr is not None:
        mask = data >= minsnr * noise
    else:
        mask = np.ones_like(data, dtype=bool)
    data = data[mask]

    if log:
        data = np.log10(data)
    img[mask] = pu.minmax(data)
    return img.astype(np.float32)
    

def radio_preprocess(
    idx, sample, 
    path="images", 
    img_size=(150, 150), 
    **kwargs
):
    """Preprocess a VLASS image."""
    try:
        radio_comp = sample.iloc[idx]
        radio_file = radio_comp["filename"]
        radio_file = os.path.join(path, radio_file)
        radio_hdu = load_radio_fits(radio_file)
        radio_data = radio_hdu.data
        return idx, scale_data(radio_data, **kwargs)
    
    except Exception as e:
        print(f"Failed on index {idx}: {e}")
        return None


def run_prepro_seq(
    sample, outfile, 
    shape=(150, 150), 
    **kwargs
):
    """Non-parallelized preprocessing for all VLASS 
    images."""
    with pu.ImageWriter(
        outfile,
        0, 
        shape, 
        clobber=True
    ) as pk_img:
        for idx in tqdm(sample.index):
            out = radio_preprocess(
                idx, 
                sample, 
                img_size=shape, 
                **kwargs
            )
            if out is not None:
                pk_img.add(out[1], attributes=out[0])
                
imbin_file = "IMG_catalog.bin"
run_prepro_seq(
    sample, 
    imbin_file, 
    shape=(150, 150), 
    path="cutouts", 
    log=True, 
    minsnr=2
)
\end{verbatim}}

The SOM can now be trained. For this experiment we used a $10\times 10$ SOM trained in three stages.

%
%
%
{\footnotesize
\begin{verbatim}
WIDTH=10
HEIGHT=10
SUF=\_h"$HEIGHT"\_w"$WIDTH"\_vlass
IMG='IMG_catalog.bin'

OUT1="SOM_B1$SUF.bin"
Pink --train $IMG $OUT1 --init random --numthreads 8 -n 
180 -p 10 \
     --som-width $WIDTH --som-height $HEIGHT \
     --dist-func unitygaussian 2.5 0.1 --inter-store 
keep \
     --euclidean-distance-shape circular  --num-iter 5

OUT2="SOM_B2$SUF.bin"
Pink --train $IMG $OUT2 --init $OUT1 --numthreads 8 -n 
180 -p 10 \
     --som-width $WIDTH --som-height $HEIGHT \
     --dist-func unitygaussian 1.5 0.05 --inter-store 
keep \
     --euclidean-distance-shape circular --num-iter 5

OUT3="SOM_B3$SUF.bin"
Pink --train $IMG $OUT3 --init $OUT2 --numthreads 8 -n 
360 -p 10 \
     --som-width $WIDTH --som-height $HEIGHT \
     --dist-func unitygaussian 0.7 0.05  --inter-store 
keep \
     --euclidean-distance-shape circular --num-iter 10
\end{verbatim}}

\begin{figure}
    \centering
    \includegraphics[width=0.99\textwidth]{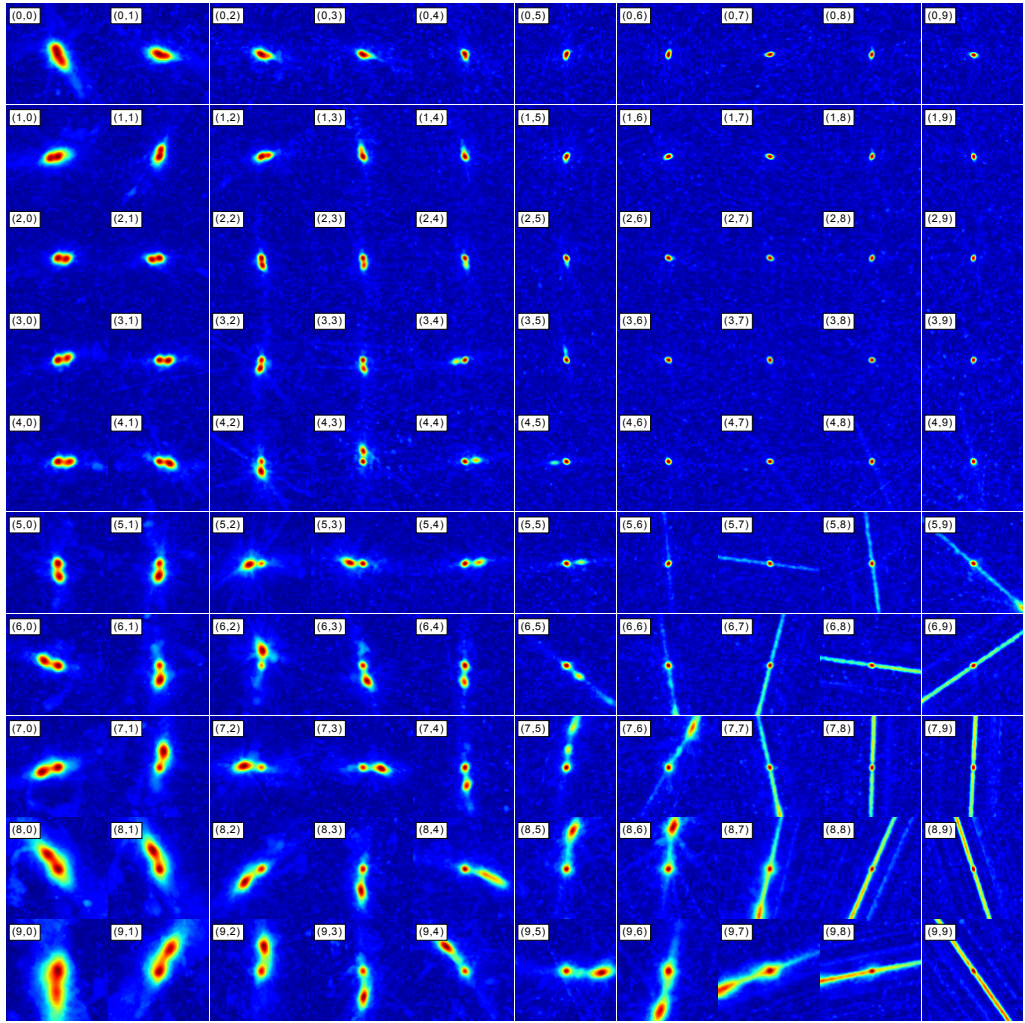}
    \caption{The $10\times 10$ SOM trained in order to estimate the sidelobe probability for each VLASS component. Each neuron shows the average of the images in that neuron. The neurons in the lower right are dominated by sidelobes. Any source that is assigned to those neurons is very likely a sidelobe.}
    \label{fig:som}
\end{figure}

\begin{figure}
    \centering
    \includegraphics[width=0.9\textwidth]{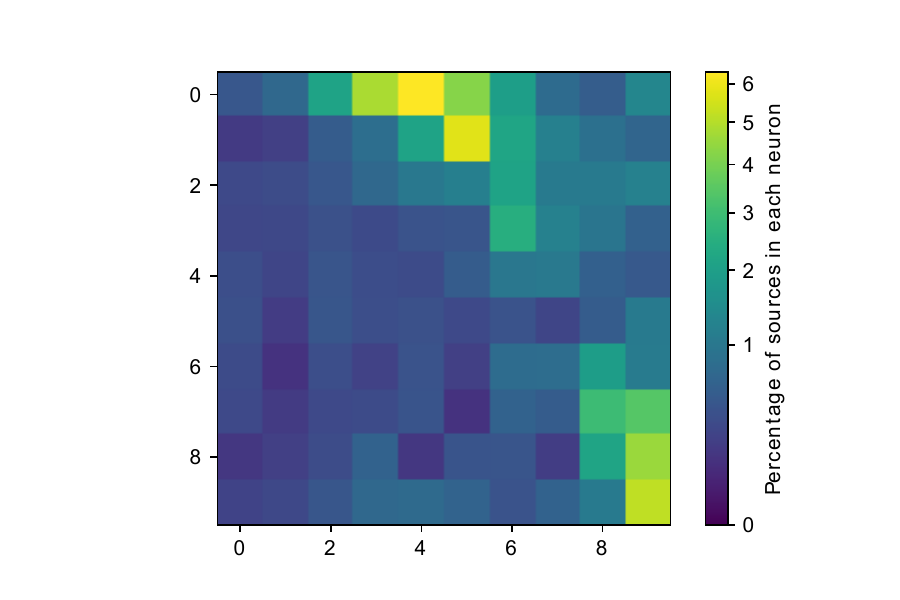}
    \caption{The distribution of sources across all the neurons.}
    \label{fig:som2}
\end{figure}

\begin{figure}
    \centering
    \includegraphics[width=15cm]{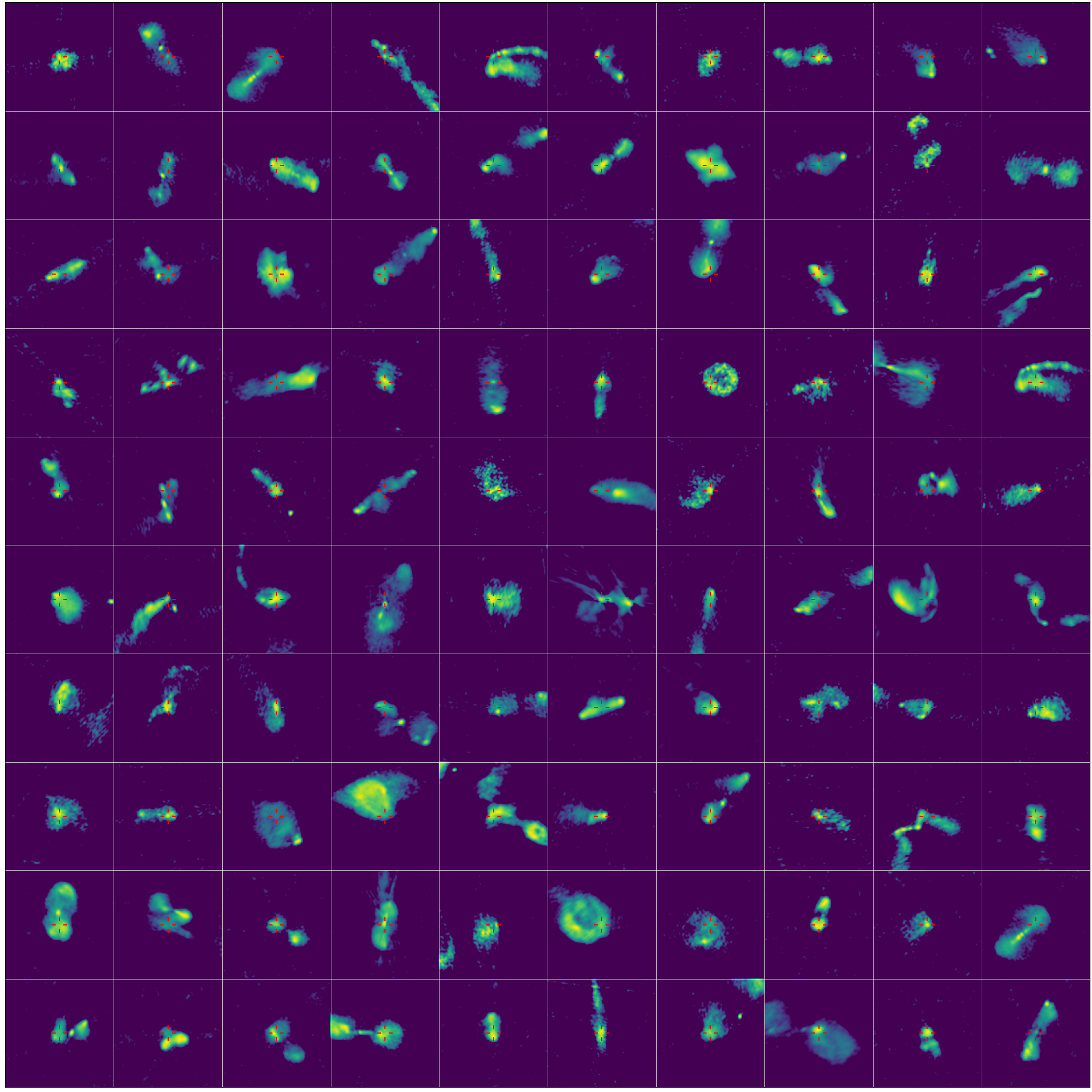}
    \caption{The sample of 100 images for neuron (9,0) that was inspected for spurious components by some of the present authors. The red cross-hair indicates the component in question. All images have been processed for use in the SOM. The median estimated sidelobe probability for this neuron is 0\%. This refers specifically to the components indicated by the red cross. Images of all 100 neurons are available at 
https://www.canfar.net/storage/vault/list/cirada/tutorials/neuronspng
    }
    \label{fig:sidelobe_10}
\end{figure}

\begin{figure}
    \centering
    \includegraphics[width=15cm]{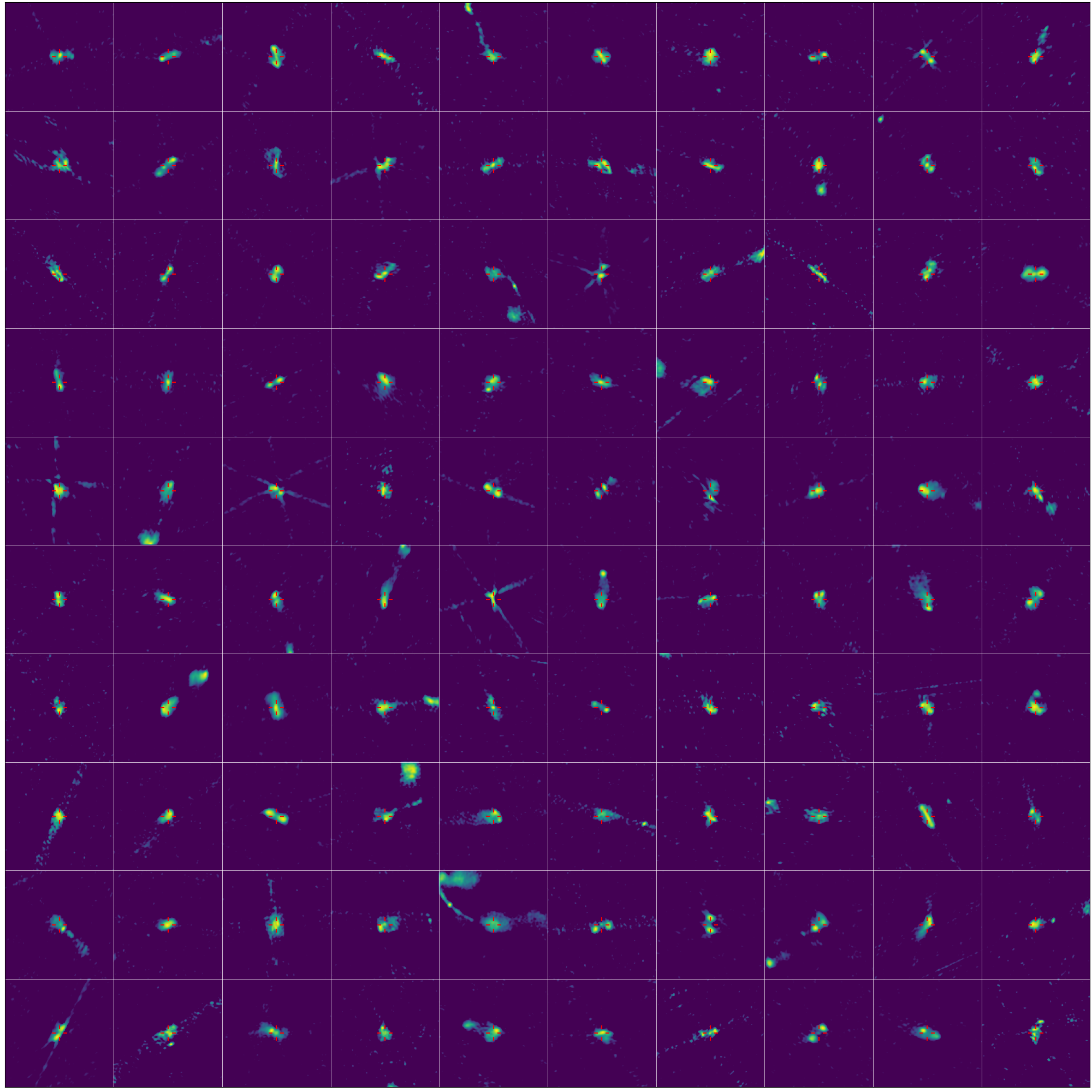}
    \caption{The sample of 100 images for neuron (0,0) that was inspected for spurious components by some of the present authors. The red cross-hair indicates the component in question. All images have been processed for use in the SOM. The median estimated sidelobe probability for this neuron is 0\%. This refers specifically to the components indicated by the red cross. Images of all 100 neurons are available at 
https://www.canfar.net/storage/vault/list/cirada/tutorials/neuronspng
    }
    \label{fig:sidelobe_7}
\end{figure}

This training process results in the SOM shown in Fig. \ref{fig:som}. The SOM has gathered together sources with similar morphology in each neuron. It can be seen in Fig. \ref{fig:som} that the neurons on the lower right are dominated by sidelobes. Figure \ref{fig:som2} shows the distribution of sources across the neurons.  Given that the SOM was trained to find sidelobes it not surprising that the lower right neurons which contain sidelobes are heavily populated.  We note that the top center neurons are also heavily populated because they contain compact sources which are the dominant morphology found in VLASS.  However, even though the SOM was trained to find sidelobes it also grouped together other types of radio morphologies.  Fig. \ref{fig:sidelobe_10} shows 100 sources in neuron (9,0). This shows that the SOM has placed extended sources with interesting morphology in this neuron. Figure \ref{fig:sidelobe_7} show 100 sources grouped together in neuron (0,0) which contain mostly double sources and a few tailed sources \citep{ODea2023}.  
Note that images of all 100 neurons are available at 
https://www.canfar.net/storage/vault/list/cirada/tutorials/neuronspng.
By choosing sources in particular neurons, samples of sources can be generated which have a certain type of morphology. 

In order to inspect the SOM and perform quality assurance one should now map the training sample onto the SOM. Here the Mapping and Transform binaries are output files. The same mapping step will also be performed on the full catalog after the user is satisfied that the SOM is well-trained and the annotations have been created.
{\footnotesize
\begin{verbatim}
Pink --map IMG_catalog.bin MAP_catalog.bin 
SOM_B3_h10_w10_vlass.bin \
    --numthreads 8 --som-width 10 --som-height 10 \
    --store-rot-flip TRANSFORM_catalog.bin \
    --euclidean-distance-shape circular -n 360 | tee 
mapping_step.log
\end{verbatim}}

Here we demonstrate how certain \textsc{pyink} functions can be used. First initialize the 
%
%
%
{\footnotesize
\begin{verbatim}
from matplotlib import pyplot as plt

# Load the training catalog.
sample = pd.read_csv("catalog.csv")

# Load the ImageBinary. Remove entries from the catalog 
# that did not preprocess correctly
imgs = pu.ImageReader("IMG_catalog.bin")
sample = sample.iloc[imgs.records].reset_index()

# Create a SOMSet with the SOM, Mapping, and Transform 
# binaries.
somset = pu.SOMSet("SOM_B3_h10_w10_vlass.bin", 
                   "MAP_catalog.bin", 
                   "TRANSFORM_catalog.bin")
\end{verbatim}}

%
%
%
{\footnotesize
\begin{verbatim}
# Plot the SOM
somset.som.plot()

# Plot a single neuron
neuron = (4, 6)
somset.som.plot_neuron(neuron)

# Plot one of the images along with its best-matching 
# neuron, which is transformed and trimmed to match the 
# frame of the image. An image can be specified by setting 
# the `idx` keyword with an integer.
pu.plot_image(
    imgs, 
    somset=somset, 
    apply_transform=True, 
    show_bmu=True, 
    idx=None
)

# Plot the number of matches to each neuron.
plt.imshow(somset.mapping.bmu_counts())
plt.colorbar()
\end{verbatim}}

If one has a set of labels for each entry in the catalog, they can be mapped onto their respective neuron as follows. For this example we generate the labels randomly.
%
{\footnotesize
\begin{verbatim}
labels = ["A", "B", "C"]
random_labels = np.random.choice(
    labels, 
    somset.mapping.data.shape[0]
)
label_counts = somset.mapping.map_labels(random_labels)

fig, axes = plt.subplots(1, 3, figsize=(16,4))
for key, ax in zip(label_counts, axes):
    im = ax.imshow(label_counts[key])
    ax.set_title(f"Label: {key}")
    plt.colorbar(im, ax=ax)
\end{verbatim}}

\subsection{From the SOM to component classification}
The final step is to create the annotations. For this experiment our goal was to assign a probability that a component corresponding to a given neuron is a sidelobe. There are a variety of ways to accomplish this.
\begin{enumerate}
    \item Plot, for each neuron, a random sample of 100 images that have been matched to the neuron. Then count the number of false positives for each grid. This number is a crude estimate of the sidelobe probability for all components match to a given neuron.
    \item Create labels for a sample of images, classifying each as either real or sidelobe. Map these labels onto the SOM and then measure the fraction of sidelobes that are matched to each neuron.
    \item Using the previous sample of classifications, train another machine learning model (likely a neural network) that takes as input the Euclidean distance to all neurons for a single image and is trained to predict whether an image is real or a sidelobe.
\end{enumerate}

We chose the first of these options, with a group of five astronomers (all coauthors of this paper) each counting the number of false positive detections. The results were recorded in a spreadsheet, the mean value was determined, and then saved to a csv. These results (the best-matching neuron, Euclidean distance, and sidelobe probability) are then appended to the catalog as follows. Note that any values with a peak-to-ring ratio above 3 or $S\_Code==`E'$ are assigned a null sidelobe probability of -1.

%
%
{\footnotesize
\begin{verbatim}
bmu = somset.mapping.bmu()
sample["Best_neuron_y"] = bmu[:, 0]
sample["Best_neuron_x"] = bmu[:, 1]
sample["Neuron_dist"] = somset.mapping.bmu_ed()

# Convert the table of probabilities into a 2D array
neuron_table = pd.read_csv("neuron_info.csv")
Psidelobe = -np.ones(
    (neuron_table.bmu_y.max() + 1, 
    neuron_table.bmu_x.max() + 1)
)
Psidelobe[neuron_table.bmu_y, neuron_table.bmu_x] = 
    neuron_table.P_sidelobe

sample["P_sidelobe"] = -np.ones(len(sample)) # Null is -1
lowPtR = 
    (sample.Peak_to_ring < 3) & (sample.S_Code != "E")
sample.loc[lowPtR, "P_sidelobe"] = 
    0.01 * Psidelobe[bmu[:, 0], bmu[:, 1]][lowPtR]
\end{verbatim}}

When one or more components fail preprocessing, the catalog at this stage of the processing will be shorter than the original. To recover the original catalog it must be loaded once more before merging in the new columns.

%
{\footnotesize
\begin{verbatim}
# First trim the table to just the new columns and a 
# unique component identifier. This saves on computing 
# time and makes the join cleaner.
neuron_cols = [
    "Best_neuron_y", 
    "Best_neuron_x", 
    "Neuron_dist", 
    "P_sidelobe"
]
update_cols = sample[["Component_name"] + neuron_cols]

# Use a left join to merge the columns
original_cat = pd.read_csv("catalog.csv")
final_cat = pd.merge(
    original_cat, 
    update_cols, 
    how="left"
)
\end{verbatim}}

\subsection{Results of the Sidelobe Study}

All components in the VLASS QL1 catalog have been matched to a neuron, but only those with
\textit{Peak\_to\_ring} $< 3$ are considered as potential sidelobes. The updated catalog features new
columns identifying, for each of the $\sim 3$  million components, its best-matching neuron in the self-organizing map (\textit{Best\_neuron\_y} and \textit{Best\_neuron\_x}), the similarity measure to its
best-matching neuron (\textit{Neuron\_dist}), and the probability that it is a sidelobe (\textit{P\_sidelobe}).

\begin{figure}
    \centering
    \includegraphics[width=0.99\textwidth]
    {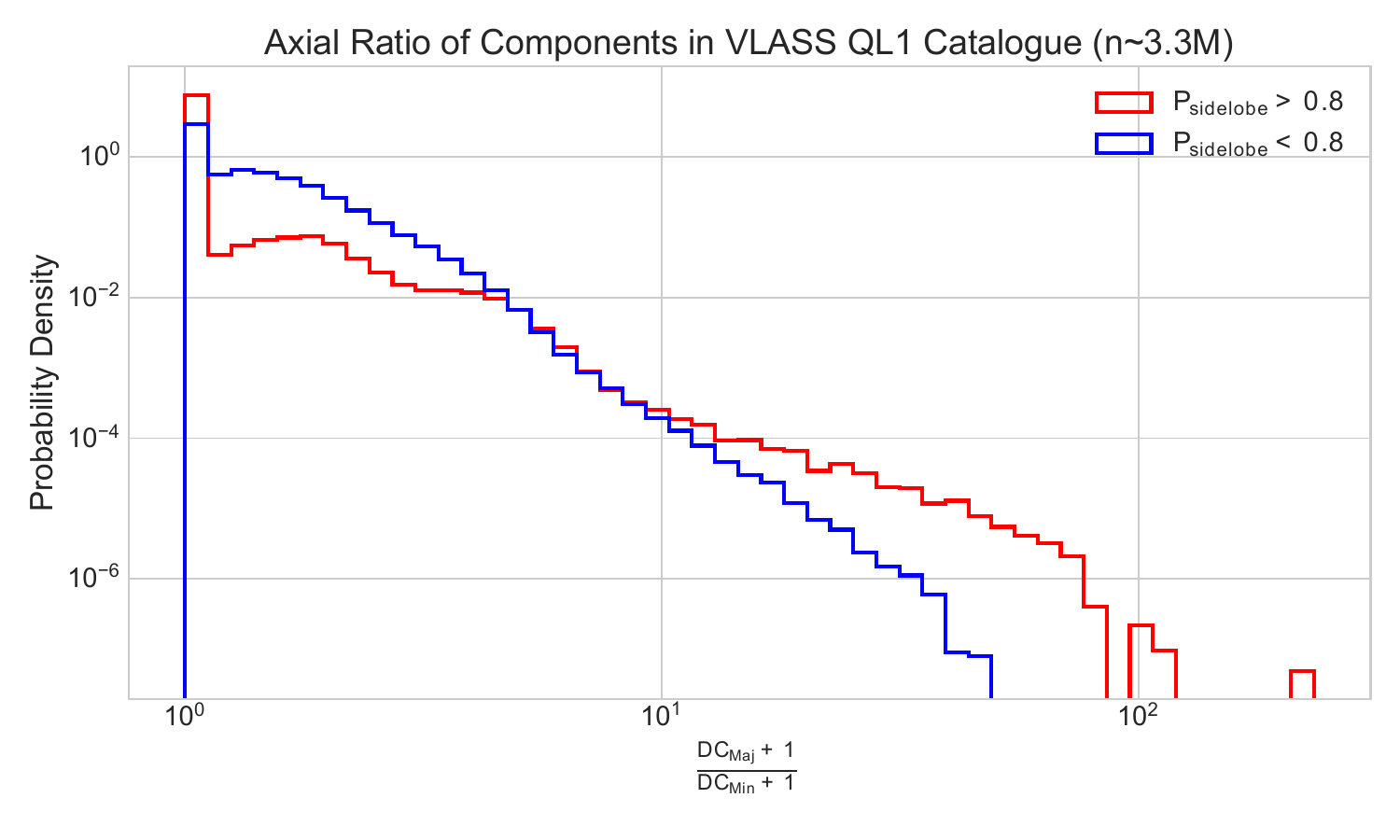}
    \caption{The axial ratio of components normalized to represent probabilities in the VLASS QL1 catalog. Catalog components with a high probability of being a sidelobe ($P > 0.8$) are shown in red. Components with a low probability of being a sidelobe ($P< 0.2$) are shown in blue. 
    }
    \label{fig:axial}
\end{figure}

 Here we show the distribution of the axial ratios for the catalog components which are very likely sidelobes ($P > 0.8$) and those which are unlikely to be sidelobes ($P< 0.2$) (Figure \ref{fig:axial}). We see that the sources with high probability of being a sidelobe tend to have a high axial ratio, i.e., they are long and narrow, as expected if they are associated with a sidelobe. This case study shows the benefit of using the SOM to identify spurious detections in limited quality data, and can thus improve the scientific usabilty of that data.

\section{Lessons Learned}
\label{sec:lessons}

\subsection{Imbalanced classes}

As the SOM is trained, the neurons are updated sequentially for each image in the training sample. First the image is matched to each neuron, the best-matching neuron is identified, and then all neurons in its vicinity are made to look more like the image. Consequentially, if a single type of morphology (i.e., one class) dominates the training sample, it will spread throughout the SOM so that it occupies a large fraction of the neurons. A rare class, on the other hand, may be relegated to very few neurons, or may even be combined with another class into a single neuron.

This issue can be addressed when selecting the training sample. Choosing a comparable number of images from each class should provide a well-balanced SOM. Morphological labels, however, are not always available. Instead, measurements based on each component or image can be used to inform the selection. For example, if one wishes to train a SOM on complex morphologies, point sources can be excluded by only using components with one or more other components within a certain radius.

If the sample selection has not resolved the issue of imbalanced classes, then a larger SOM (i.e., the user can chose a larger number of neurons) may offer an improvement. This will provide the space needed for the dominant classes to spread throughout the SOM, while the rarer classes will still be represented in a few neurons.

A more complex approach is to use multiple SOMs to create the final hierarchical model. The first SOM is used as a coarse morphological classifier. The user can then select the neurons that fit their scientific interest and train a second SOM on the images that match to them. This is a more computationally expensive approach, but also the most powerful. 

{ An alternative method of addressing imbalanced classes is the Synthetic Minority Oversampling Technique (SMOTE,
\citep{Chawla2011}).}

\subsection{Weighting of multiple channels}
\label{sec:multi-chan_weight}
The relative weights of each channel in the preprocessed images will affect the patterns identified by the SOM. Consider a 2-channel image stack containing the radio and IR wavelengths. Since the surface number density of detected IR sources is significantly
higher than that of detected radio sources, an IR weighting that is too high will result in neurons whose main feature is the separation between two independent galaxies. Finding the appropriate weights is critical in ensuring that as few neurons as possible are devoted to irrelevant morphological features.

A reasonable starting place is to set the weights such that the sum of all pixels in each image has the same median. First process each image with a weight of 1, sum all of the pixels in each image, and measure the median for the ensemble of images. The \verb|ImageReader.reweight| function in \textsc{pyink} can then be used to adjust the weights based on this measurement. Further experimentation is necessary to determine if this weighting is appropriate for the science goal.

In the above example with both a radio and IR channel, the presence of background galaxies complicates the choice of weights. Summing together all pixels in the IR channel results in large values that are not indicative of the relevant structure. Normalizing by the medians therefore under-represents the important IR structures.

Addressing this issue is not a trivial task. Several strategies can be tried, but none is a clear solution to the problem. The overall goal is to suppress the irrelevant structures as much as possible, but it is not always possible to tell which sources can be excluded. One approach is to use a simple channel (the radio channel for this example) to mask out structures in the complex one. For example, a convex hull created from the preprocessed radio channel can be applied to the IR channel, excising all galaxies outside of the radio structures. The channel weights can then be determined from the sum of all pixels, as before. Alternatively, or additionally, the image can be multiplied by a radial function (e.g. a gaussian) to down-weight any non-central sources. This, however, assumes that the relevant structures are always central, which is often not correct.

Overall, deciding on the channel weights comes down to experimentation. It is difficult to optimize the weights \textit{a priori}, and more advanced processing may be important in making the most of the data.

\subsection{Scale Invariance}

While the SOM trained using \textsc{PINK} is invariant to both rotations and flips, it is not scale invariant. This presents a problem for astronomical images, where two sources with the same morphology can be drastically different in angular size. { If redshifts/distances were available for all sources in the sample, which is not true here, or in general, then transforming to a physical scale would reduce the range in image sizes.}

When a SOM is trained on a collection of similar morphologies with varying sizes, any non-central emission is smeared in the resulting neurons. Consider the simple example of a radio double. Each preprocessed image is centered on one of the radio components. SOM training effectively stacks these images. The central component is added coherently across all images. The non-central component is aligned to a common angle, but the varying separations result in blended emission in the neuron.

A coarse SOM -- one with few neurons -- exacerbates this effect. The central component dominates the total brightness, while the superposition of non-central components at a range of separations produces a long, faint trail extending from the center. Similar morphologies will be grouped together, but the neurons lose meaning.

The only robust way to address this issue is to increase the number of neurons in the SOM. With a smaller range of separations that is matched to a neuron, the position of the non-central component is better represented in the trained neuron. Instead of forming a long, faint trail, the SOM will more closely resemble the preprocessed images, though some smearing will still exist. Unfortunately this can quickly become computationally expensive, increasing both the training/mapping time and the required GPU memory (see Equation \ref{eqn:mem}). { Note that because we have used a sufficient number of neurons (100), the SOM has grouped together into neighborhoods sources with similar angular sizes (e.g., Figures \ref{fig:som}, \ref{fig:sidelobe_10}, \ref{fig:sidelobe_7}). So in practice, the SOM is not limited to populations of objects at similar distances.}

\subsection{Combining with a classification algorithm}

When using a SOM on its own, the information that can be inferred for a component is determined only from its best-matching neuron. However, the mapping file records the similarity (Euclidean distance) between an image and every neuron, not just the best-matching neuron. These similarity measurements can be combined with another machine learning algorithm in order to create a full classification model. The SOM handles the image analysis, creating weights (the similarities) to each of the prototype classes (the neurons), while the other machine learning model performs the final classification. This approach requires a labelled sample to be used to train the classifier. The classifier can, in principle, be any supervised machine learning algorithm, such as a neural network. { Algorithms such as k-nearest neighbors \citep{fix1951}, convolutional neural network \citep{Oshea2015}, or random forest \citep{Ho1995} are potentially useful.}

\section{Summary}

The self-organizing map (SOM) is an unsupervised machine learning algorithm that projects a many-dimensional dataset onto a two- or three-dimensional lattice of neurons. This dimensionality reduction allows the user to better visualize common features of the data and develop algorithms that would not be feasible
to apply to the initial, much larger datasets. 

The SOM implementation called PINK \citep{PINK} incorporates rotation and flipping invariance so that the SOM algorithm may be applied to astronomical images. In this cookbook we provide instructions for working with PINK, including preprocessing the input images, training the model, and offering lessons learned through experimentation. The problem of imbalanced classes can be improved by careful selection of the training sample and increasing the number of neurons in the SOM. Because PINK is not scale-invariant, structure can be smeared in the neurons. This can be improved by increasing the number of neurons in the SOM. 

We also introduce \textsc{pyink}, a Python package used to read and write PINK binary files, assist in common preprocessing operations, perform standard analyses, visualize the SOM and preprocessed images, and create image-based annotations using a graphical interface. A tutorial is also provided to guide the user through the entire process. We show that PINK is generally able to group VLASS radio sources with similar morphology together. 
PINK has been applied to the VLASS Quick Look epoch 1 data to produce a probability for each source in the VLASS QL1 catalog \citep{Gordon2021} that the source is actually a sidelobe.  All software discussed here is available at \url{https://cirada.ca/vlasspipeline#pipeline3}.

%




\section*{Acknowledgements}
CO, SB, AV and MB are supported by NSERC -- the National Science and Engineering Research Council of Canada. Y.A.G. is supported by U.S. National Science Foundation grant AST 20-09441.
Partial support for the contributions from LR come from US National Science Foundation grant AST17-14205 to the University of Minnesota.

The Canadian Initiative for Radio Astronomy Data Analysis (CIRADA) is funded by a grant from the Canadian Foundation for Innovation (CFI).
This work makes use of data provided by the National Radio Astronomy Observatory, a facility of the National Science Foundation operated under cooperative agreement by Associated Universities, Inc.
This research used the facilities of the Canadian Astronomy Data Centre operated by the National Research Council of Canada with the support of the Canadian Space Agency.




\bibliographystyle{elsarticle-harv} 






\end{document}